\documentclass{jfm}

\usepackage{natbib}
\usepackage{graphicx}
\usepackage{subfig}
\usepackage{amssymb,amsmath}
\usepackage{mathrsfs}

\ifCUPmtlplainloaded \else
  \checkfont{eurm10}
  \iffontfound
    \IfFileExists{upmath.sty}
      {\typeout{^^JFound AMS Euler Roman fonts on the system,
                   using the 'upmath' package.^^J}%
       \usepackage{upmath}}
      {\typeout{^^JFound AMS Euler Roman fonts on the system, but you
                   dont seem to have the}%
       \typeout{'upmath' package installed. JFM.cls can take advantage
                 of these fonts,^^Jif you use 'upmath' package.^^J}%
      }
  \else
  \fi
\fi

\ifCUPmtlplainloaded \else
  \checkfont{msam10}
  \iffontfound
    \IfFileExists{amssymb.sty}
      {\typeout{^^JFound AMS Symbol fonts on the system, using the
                'amssymb' package.^^J}%
       \usepackage{amssymb}%
         \let\leq=\leqslant
         
      }{}
  \fi
\fi

\ifCUPmtlplainloaded \else
  \IfFileExists{amsbsy.sty}
    {\typeout{^^JFound the 'amsbsy' package on the system, using it.^^J}%
     \usepackage{amsbsy}}
    {}
\fi

\makeatletter
\gdef\@underjournal{%
  \vbox to 6\p@{\noindent
    \parbox[t]{3in}{\normalfont\indexsize{\itshape Accepted
      for publication in J.\ Fluid\ Mech. This is a preprint version, subject to further typesetting and layout. \copyright Cambridge University Press
      }\\[2.5\p@]
      {\ \ }}%
  }%
}
\makeatother

\newcommand{\Pd}[2][]{\displaystyle \frac{\partial #1}{\partial #2} }

\newcommand{\Pdt}[2][]{\textstyle \partial #1/\partial #2 }
\newcommand{\Odt}[2][]{\textstyle {{\rm d} #1}/{{\rm d} #2} }
\newcommand{\f}[1]{\overline{#1}}
\newcommand{\ft}[1]{\widetilde{#1}}

\renewcommand{\vec}[1]{\mbox{\boldmath $ #1 $}}

\title[Momentum and energy transport tubes]{Flow visualization using momentum and energy transport tubes and applications to turbulent flow in wind farms}

\author[Johan Meyers and Charles Meneveau]%
{J\ls O\ls H\ls A\ls N\ns M\ls E\ls Y\ls E\ls R\ls S$^1$\thanks{Email address for correspondence: johan.meyers@mech.kuleuven.be} \and C\ls H\ls A\ls R\ls L\ls E\ls S\ns M\ls E\ls N\ls E\ls V\ls E\ls A\ls U$^2$}

\affiliation{$^1$Department of Mechanical Engineering, KU Leuven,\\ Celestijnenlaan~300A, B3001~Leuven, Belgium\\[\affilskip]
$^2$Department of Mechanical Engineering \& Ctr. for Environmental and Applied Fluid Mechanics, Johns Hopkins University, 3400~N.~Charles~Street, Baltimore~MD~21218, USA}

\date{14 October 2012}
\begin{document}

\maketitle

\begin{abstract}
As a generalization of the mass-flux based classical stream-tube, the concept of momentum and energy transport tubes is discussed as a flow visualization tool. These transport tubes have the property, respectively, that no fluxes of momentum or energy exist over their respective tube mantles. As an example application using data from large-eddy simulation, such tubes are visualized for the mean-flow structure of turbulent flow in large wind farms, in fully developed wind-turbine-array boundary layers. The three-dimensional organization of energy transport tubes changes considerably when turbine spacings are varied, enabling the visualization of the path taken by the kinetic energy flux that is ultimately available at any given turbine within the array.
\end{abstract}

\section{Introduction}
The notion of a stream-tube \citep{Batchelor1967,Fay94} as a tool for flow analysis and visualization is particularly useful because it maintains constant volume or mass flux across sections. A popular example of stream-tube based dynamical analysis is the actuator disk and ideal flow model of wind turbines, establishing the relationship between power extraction and fluxes of kinetic energy at the stream-tube inlet and outlet (see \citealp{burton2001}).  The classical analysis is valid for steady and ideal flow. However, in three-dimensional high-Reynolds number turbulent flows, the transport of mean momentum and mean-flow kinetic energy is often dominated by Reynolds stresses. Then the averaged flux of mass being visualized by a stream tube is no longer representative of the transport of other flow properties such as momentum or energy. As a generalization of the classic stream-tube, we consider the concept of momentum and energy transport tubes,  defined such that there is no average transport of  momentum or mean-flow total mechanical energy over their respective tube mantles. These tubes enable us to visualize the trajectory of transported properties across the flow. We illustrate the approach for the case of turbulent flow through large wind farms, looking for an answer to the question: ``Where does the kinetic energy come from that is ultimately extracted by a given wind turbine?'' To this end, we employ data obtained from large-eddy simulations (LES) of fully developed wind-turbine-array boundary layers \citep*[following][]{Calaf2010}, with different stream-wise and span-wise turbine spacings and wind turbine arrangements.

In Section~\ref{s:massmomenergy} the concept of transport tubes for mean-flow momentum and mean-flow mechanical energy is presented. Next, to illustrate the concept, some analytically tractable laminar flow examples are provided in Section~\ref{s:laminar}. Subsequently, in Section~\ref{s:windfarm} we demonstrate the use of these tubes for the interpretation of momentum and energy fluxes in large wind farms. There, we first evaluate conventional averaged-flow stream tubes that pass through a target wind turbine disk and then investigate momentum and energy transport tubes for various turbine spacings in the wind farm. Further discussion is provided in Section~\ref{s:discussion}. Conclusions are presented in Section~\ref{s:conclusions}.

\section{Mass, momentum, and energy tubes}\label{s:massmomenergy}
We focus on incompressible viscous, statistically stationary turbulent flows with constant density $\rho$. A classical streamline of a stationary mean flow field is commonly defined as a curve $\Gamma$ parametrized by  $\vec{x}(s)\in \Gamma$ ($s\in \mathbb{R}$), for which $\vec{x}(s)\times \f{\vec{u}}=0$, with $\f{\vec{u}}=[\f{u}_1, \f{u}_2, \f{u}_3]$ the mean velocity vector in a particular frame of reference. A stream tube is then constructed by selecting a closed curve $C$, which is not anywhere tangent to the velocity, and considering the bundle of all streamlines through that curve $C$ \citep{Batchelor1967}. Consider a volume of stream tube $\Omega$, bound by the tube mantle $M$, and two cross sections $A_1$ and $A_2$. The volume-integrated continuity equation leads to $\iint_{A_2} \rho \f{u}_i n_i \ {\rm d}\vec{x} + \iint_{A_1} \rho \f{u}_i n_i \ {\rm d}\vec{x} =  0$, where ${\bf n}$ is the outward directed normal to the stream-tube control volume. No mass flows through the tube mantle $M$  since by construction $\f{u}_i n_i=0$ there.

To construct momentum or energy transport tubes, which have the property that there is on average no exchange of momentum or energy through the corresponding tube's mantle, we consider the vector fields formed by the total flux of these quantities. The total flux includes advective, turbulent and viscous fluxes.
For  the transport of linear momentum, we consider a direction characterized by constant unit vector $\vec{\zeta}$ and components $\zeta_i$ (as an example,
$\vec{\zeta}$ could be any one of the Cartesian unit vectors $\vec{i}$, $\vec{j}$ or $\vec{k}$).
Hence, the $\vec{\zeta}$-momentum is $\rho u_i \zeta_i$, and for statistically steady flow its transport equation is given by
\begin{equation}
\frac{\partial} {\partial x_j} (\rho \f{F}_{m,j})= -\Pd[\f{p}]{x_i} \zeta_i  + \f{f}_i \zeta_i,
\end{equation}
where $\f{f}_i$ represents the  body force, and
\begin{equation}
 \f{F}_{m,j} =  \f{u}_j ~( \f{u}_i \zeta_i) ~+  ~\left(\f{u'_i u'_j} - 2\nu\f{S}_{ij} \right)\zeta_i
 \label{eq:defomflux}
\end{equation}
is the flux vector field of linear momentum (per unit mass) in the $\vec{\zeta}$ direction (the index ``$m$'' refers to momentum), $\nu$ is the kinematic fluid viscosity ($\nu=\mu/\rho$ with $\mu$ the dynamic viscosity), and $\f{S}_{ij} = (\Pdt[\f{u}_i]{x_j} + \Pdt[\f{u}_j]{x_i})/2$ the mean rate-of-strain tensor. Also, a linear-momentum-transporting velocity field $\f{\vec{u}}_{m}$ can be defined according to:
\begin{equation}
\f{u}_{m,j} = \f{u}_j + \frac{[\f{u'_i u'_j}-2\nu \f{S}_{ij}]\zeta_i}{\f{u}_k \zeta_k}.
\label{eq:defum}
\end{equation}
A similar notion of a diffusion velocity has  been introduced in the context of deterministic particle transport methods \citep{Hermeline89}, and later used in particle vortex methods for viscous flows \citep{DegondMustieles90,GrantMarshall05}. Some similarity also exists with the notion of Favre averaging for compressible flow, where a mass-transport velocity field is defined by dividing the mass flux by the average density \cite{Favre1976,Smits2006}.

Constructing a tube based on $\f{\vec{F}}_m$ (or $\f{\vec{u}}_m$), we now find
\begin{equation}
\iint_{A_2} \rho   \f{F}_{m,j}  n_j \ {\rm d}\vec{x}~ + \iint_{A_1} \rho   \f{F}_{m,j} n_j \ {\rm d}\vec{x} = -\iiint_\Omega \Pd[\f{p} \zeta_i]{x_i} \ {\rm d}\vec{x}~ + \iiint_\Omega  \f{f}_i \zeta_i \ {\rm d}\vec{x}.
\end{equation}
No momentum is transported through the tube mantle, since by construction $ \f{F}_{m,i} n_i=0$ there. As a result, on the tube's cross-sections $A_1$ and $A_2$, the flux of linear momentum is constant, except for integral effects of sources and sinks of momentum, $\zeta_i\Pdt[\f{p}]{x_i} $, and $\f{f}_i \zeta_i$ in the tube. The pressure effects can also be written in terms of the pressure at the inlet, outlet and mantle using
$-\iiint_\Omega \partial(\f{p} \zeta_i)/\partial{x_i} \ {\rm d}\vec{x} =  \iint_{A_1} \f{p} ~\zeta_i  n_i \ {\rm d}\vec{x} + \iint_{A_2} \f{p} ~\zeta_i  n_i \ {\rm d}\vec{x} + \iint_{M} \f{p} ~\zeta_i  n_i \ {\rm d}\vec{x}$. We remark that a classical jet evolving at constant pressure may be considered a stream-wise momentum transport tube, since the momentum flux across its sections remains constant and no forces or momentum fluxes act on its mantle even though a mass flux crosses the mantle -- see Section~\ref{s:laminar} for details, where an elaboration of momentum and energy tubes (the latter is defined below) is presented for some simple canonical laminar-flow cases. Note that the momentum-flux vector and related tube geometry depend directly on the choice of the direction $\vec{\zeta}$ in which linear momentum is defined. The tube geometry also depends upon the velocity of the reference frame. Like classic stream tubes, generalized transport tubes are not Galilean invariant. For illustration on wind-farm cases in Section~\ref{s:windfarm}, we focus on stream-wise momentum along the incident wind direction only.

Similarly we consider mean-flow energy-transport tubes, based on the transport equation for mean-flow kinetic energy ($\rho{K} =\rho \f{u}_i\f{u}_i/2$):
\begin{equation}
\frac{\partial}{\partial x_j} (\rho \f{F}_{K,j}) = - \Pd[\f{u}_i \f{p}]{x_i} + \rho \f{u'_i u'_j} \Pd[\f{u}_i]{x_j}  -  {2} \mu \f{S}_{ij}  \f{S}_{ij} + \f{u}_i \f{f}_i,
\end{equation}
where
\begin{equation}
\f{F}_{K,j} =  K \f{u}_j+  (\f{u'_i u'_j} - 2 \nu \f{S}_{ij}) \f{u}_i ,
\label{eq:defoKflux}
\end{equation}
is the total kinetic energy flux vector field per unit mass, and the kinetic energy transport velocity  is
$ \f{u}_{K,j} = \f{F}_{K,j}/K$. These vector fields may be used to construct energy-transport tubes.

A difficulty for the interpretation of momentum and energy tubes is the fact that the pressure gradient acts as a source term. This is less of an issue when the pressure gradient is only due to an external pressure difference, such that the gradients only relate to an external force, and power inserted in the system respectively. However, when local accelerations or decelerations impact the local pressure (e.g. near wind turbines, see below), the interpretation of these sources is less natural. For the study of energy fluxes, this can be remedied by looking at mean-flow total mechanical energy tubes. To this end, we decompose the mean pressure gradient as $\nabla \f{p} = \nabla p_\infty + \nabla \hat{p} = - \vec{f}_\infty + \nabla \hat{p}$, where $\vec{f}_\infty$ may be an external driving force per unit volume. Denoting the total mean-flow mechanical energy per unit mass as
\begin{equation}
{E} = \f{u}_i\f{u}_i/2 + \hat{p}/\rho,
\end{equation}
its transport equation reads
\begin{equation}
\frac{\partial (\rho \f{F}_{E,j})}{\partial x_j} =  \rho \f{u'_i u'_j} \Pd[\f{u}_i]{x_j}  -  {2}\mu \f{S}_{ij}  \f{S}_{ij} + \f{u}_i (\f{f}_i+f_{i,\infty}),
\end{equation}
with the total mechanical energy transport vector field $\f{\vec{F}}_E$ defined according to
\begin{equation}
\f{F}_{E,j} =   E \f{u}_j+ (\f{u'_i u'_j}- 2 \nu \f{S}_{ij}) \f{u}_i.
\label{eq:defue}
\end{equation}
Related to this, a transport-velocity field may also be defined as $\f{F}_{E,j}/E$. Constructing a tube based on $\f{F}_{E,j}$, we now find
\begin{eqnarray}
\iint_{A_2} \rho  \f{F}_{E,j}  n_j \ {\rm d}\vec{x}~ + \iint_{A_1}  \rho  \f{F}_{E,j}  n_j \ {\rm d}\vec{x} &=& - \iiint_\Omega  \left( {2}\mu \f{S}_{ij}  \f{S}_{ij} -  \rho \f{u'_i u'_j} \Pd[\f{u}_i]{x_j}  \right)\ {\rm d}\vec{x}\nonumber \\ ~ & & + \iiint_\Omega  \f{u}_i (\f{f}_i+f_{i,\infty})\ {\rm d}\vec{x}.
\end{eqnarray}
No mean total mechanical energy  is transported through the tube mantle and the flux across sections of the tube is constant, except for sources/sinks of mean-flow kinetic energy by the distributed force ($\f{u}_i \f{f}_i$), by mean-flow viscous dissipation (sink), and due to production of turbulent kinetic energy $-\rho \f{u'_i u'_j} \Pdt[\f{u}_i]{x_j}$ (typically also a sink of mean energy). For conservative force-fields, one of course also has the option of  including it into the definition of $E$ via its potential function. Examples of energy tubes for some simple laminar-flow cases are briefly discussed in next section.

Finally, in the particular case of ideal (inviscid) and steady laminar flow, we have $\f{u'_iu'_j}=0$ and $2 \nu \f{S}_{ij} = 0$. Hence, it is obvious that $ \f{\vec{u}} = \f{\vec{u}}_{m} = \f{\vec{F}}_{K}/K = \f{\vec{F}}_{E}/E$, from which it follows that stream tubes, momentum transport tubes and energy transport tubes all collapse, as conventionally used in ideal-flow, stream-tube analysis. However, in  turbulent flows Reynolds stresses can affect momentum and energy fluxes considerably, so that these different tubes may differ greatly.  This is illustrated with applications to flow in wind-farm boundary layers in Section~\ref{s:windfarm}. Furthermore, we can remark that transport tubes for other quantities such as vorticity, helicity, temperature or elements of Reynolds stress may be derived accordingly.

\section{Transport tubes for some simple laminar flows}\label{s:laminar}
To first illustrate the concept of momentum and energy tubes, they are briefly elaborated in the current section for a few simple canonical laminar-flow cases.
\subsection{Couette flow}
\begin{figure}
  \begin{center}
  \includegraphics[width=0.7\textwidth]{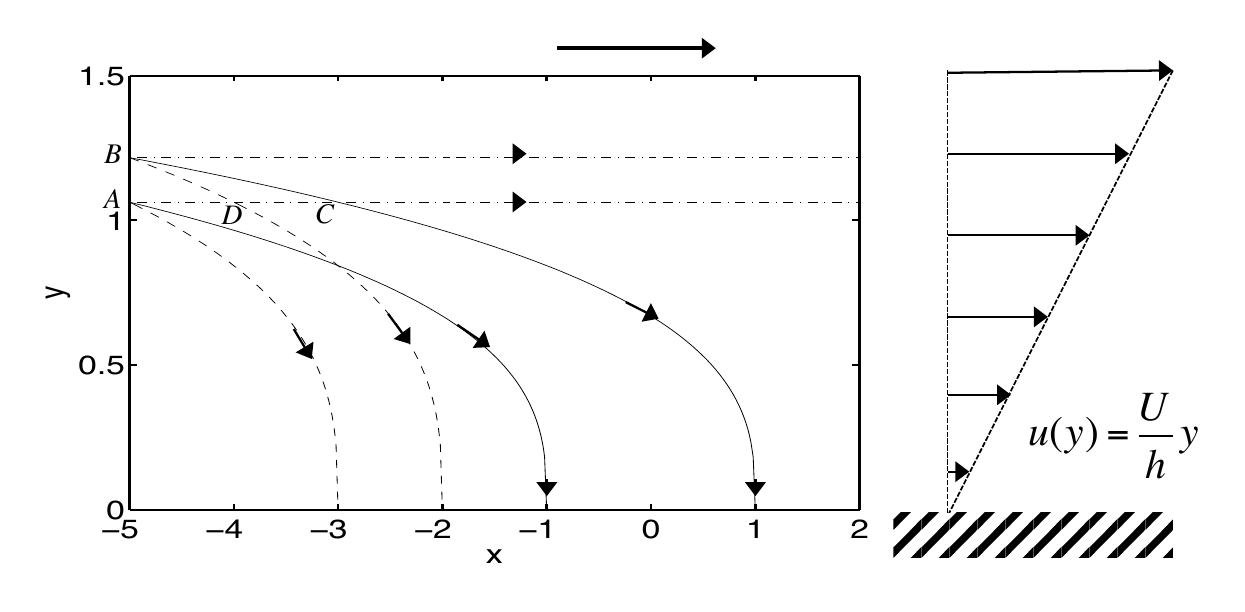}
  \end{center}
  \caption{Couette flow: dot-dashed lines show sample streamlines, solid lines are momentum  lines, while dashed lines denote kinetic energy lines.}\label{f:Couette-lines}
\end{figure}

We first consider laminar Couette flow, with an along-boundary (horizontal) velocity profile given by $u(y) = y (U/h)$ and vertical velocity $v=0$. Then the momentum flux is given by the following two components: $F_{m,1} = u u_{m} = (yU/h)^2$ and  $F_{m,2} = u v_{m}(y) = - \nu (\partial u/\partial y)  = - \nu (U/h)$. Therefore the slope of the tangent lines of this vector field is given by $dy_m/dx = F_{m,2}/F_{m,1}=v_{m}/u_{m} = -(\nu h/U) y_m^{-2} = - Re^{-1} (y_m/h)^{-2}$ (where $Re=Uh/\nu$). Integration yields momentum lines of the form
\begin{equation}
\frac{y_m(x)}{h}= \left[\frac{3}{Re}\frac{x_0-x}{h}\right]^{1/3}.
\end{equation}
Figure \ref{f:Couette-lines} shows the resulting shape of these lines for two values of $x_0/h=\pm 1$ thus enclosing a (2D) momentum transport tube. We used $Re=10$ and the lengths shown are in units of $h$. The dot-dashed lines show regular streamlines.  The momentum transport  lines can be interpreted as follows: the flux of x-direction (horizontal) momentum flux that enters at $x=-5$ (A--B) is transferred through this tube downwards towards the solid wall ($y_m=0$), where it is equalled by the viscous drag acting between $x=-1$ and $x=1$. By definition there is no net momentum flux of any type crossing the solid lines, and since the problem is steady and there are no further forces acting (e.g. pressure), the entire momentum flux is absorbed at the wall. Alternatively, one may regard (e.g.) the bottom horizontal (dot--dashed) line as the top wall that is being dragged from left to right. Then the total drag force acting on this top wall between $x=-5$ and $x=-3$ (segment A--C) is ``transmitted'' via the momentum transport tube towards the bottom wall as indicated by the solid lines. Near the wall, the transport velocity becomes vertical as more and more of the momentum transport occurs through viscous diffusion, while the momentum being transported vanishes. Because of the latter effect, the magnitude of the vertical transport velocity diverges to infinity, while the transport lines remain well-defined. We also remark that at increasing $Re$ (or increasing $h$ away from the bottom wall), the transport lines become more horizontal, as inertia in the horizontal direction dominates over viscous diffusion. Close to the bottom wall, viscous diffusion dominates.

Next, we consider the kinetic energy transport flux, see Eq.~(\ref{eq:defue}). Similarly as with momentum transport, one can show that the energy lines are given by
\begin{equation}
\frac{y_E(x)}{h}= \left[\frac{6}{Re}\frac{x_0-x}{h}\right]^{1/3}.
\end{equation}
The resulting lines, starting out at the same points as the streamlines and momentum lines at $x=-5$ in Figure~\ref{f:Couette-lines} and shown using dashes, curve down towards the wall more quickly than the momentum lines. The entire flux of kinetic energy that enters the energy tube at $x=-5$ (segment A-B) is transported towards the wall while being dissipated into heat inside the tube. Since no work is being done on the bottom wall, the entire energy is dissipated inside the tube before reaching the bottom wall. Conversely, there is work done by  a moving top wall, e.g. along segment A-D, which is then is dissipated into heat inside the energy tube.

\subsection{Poiseuille flow}

\begin{figure}
  \begin{center}
  \includegraphics[width=1\textwidth]{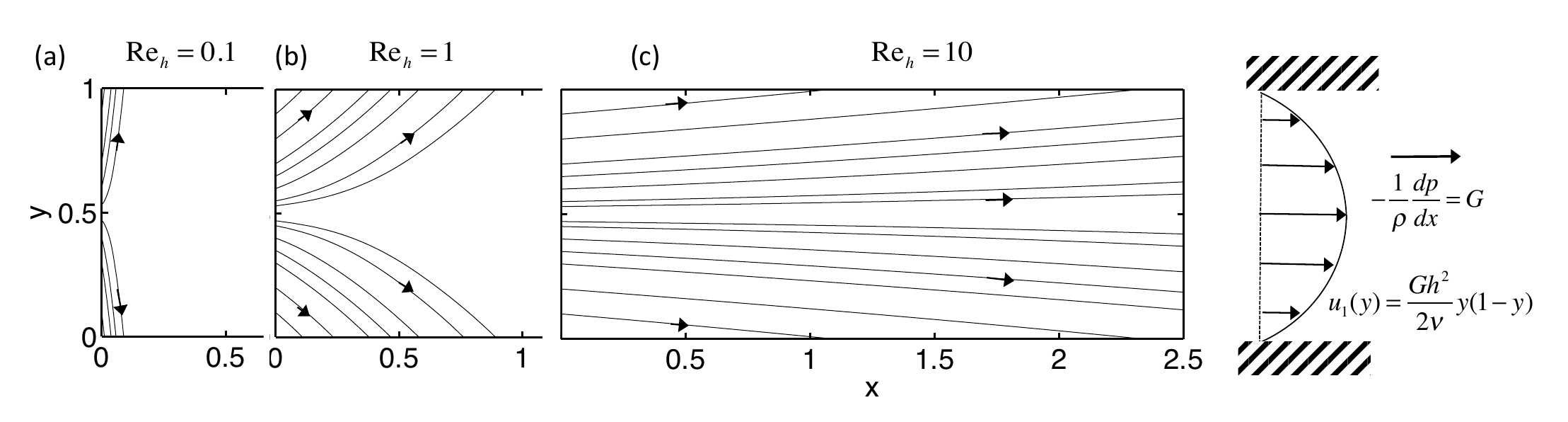}
  \end{center}
  \caption{Momentum lines for Poiseuille flow at three Reynolds numbers}\label{f:Poiseuille-lines}
\end{figure}

Similarly, if we consider simple laminar Poisseuille flow of the form $u(y) = y(1-y)Gh^2/(2\nu)$, with $G=-(1/\rho)\Odt[p]{x}$ and $y$ non-dimensionalised with channel total height $h$, we obtain $F_{m,1}=u(y)^2$ and $F_{m,2}=-(1-2y)Gh/(2\nu)$. The momentum-line slopes become $\Odt[y_m]{x}=-(8Re_h)^{-1} (1-2y)[y^2(1-y)^2]^{-1}$, where $Re_h=U h/\nu$ and $U=Gh^2/(16 \nu)$ is the channel mean velocity. The slope depends upon the Reynolds number, with steeper slopes corresponding to lower Reynolds number (stronger diffusion transport) as expected.  Integration yields momentum lines given by $y=y_m(x)$ and passing through
 $(x_0,y_0)$, according to:
 \begin{equation}
  4 \ln \left(\frac{y-\frac{1}{2}}{y_0-\frac{1}{2}}\right) + (2y-3)(2y+1)(1-2y)^2 - (2y_0-3)(2y_0+1)(1-2y_0)^2  = \frac{16}{Re_h} (x - x_0).
 \end{equation}
  The results are shown in Fig. \ref{f:Poiseuille-lines} for three Reynolds numbers.  The interpretation is that the momentum added in the bulk of the flow through the pressure gradient is transported towards the side-walls as shown in the figure. We note that kinetic energy lines have the same shape but with twice the slope, and the work done by the pressure gradient is dissipated entirely, before reaching the walls.

\subsection{Laminar round jet and wake}
\begin{figure}
  \begin{center}
  \includegraphics[width=0.5\textwidth]{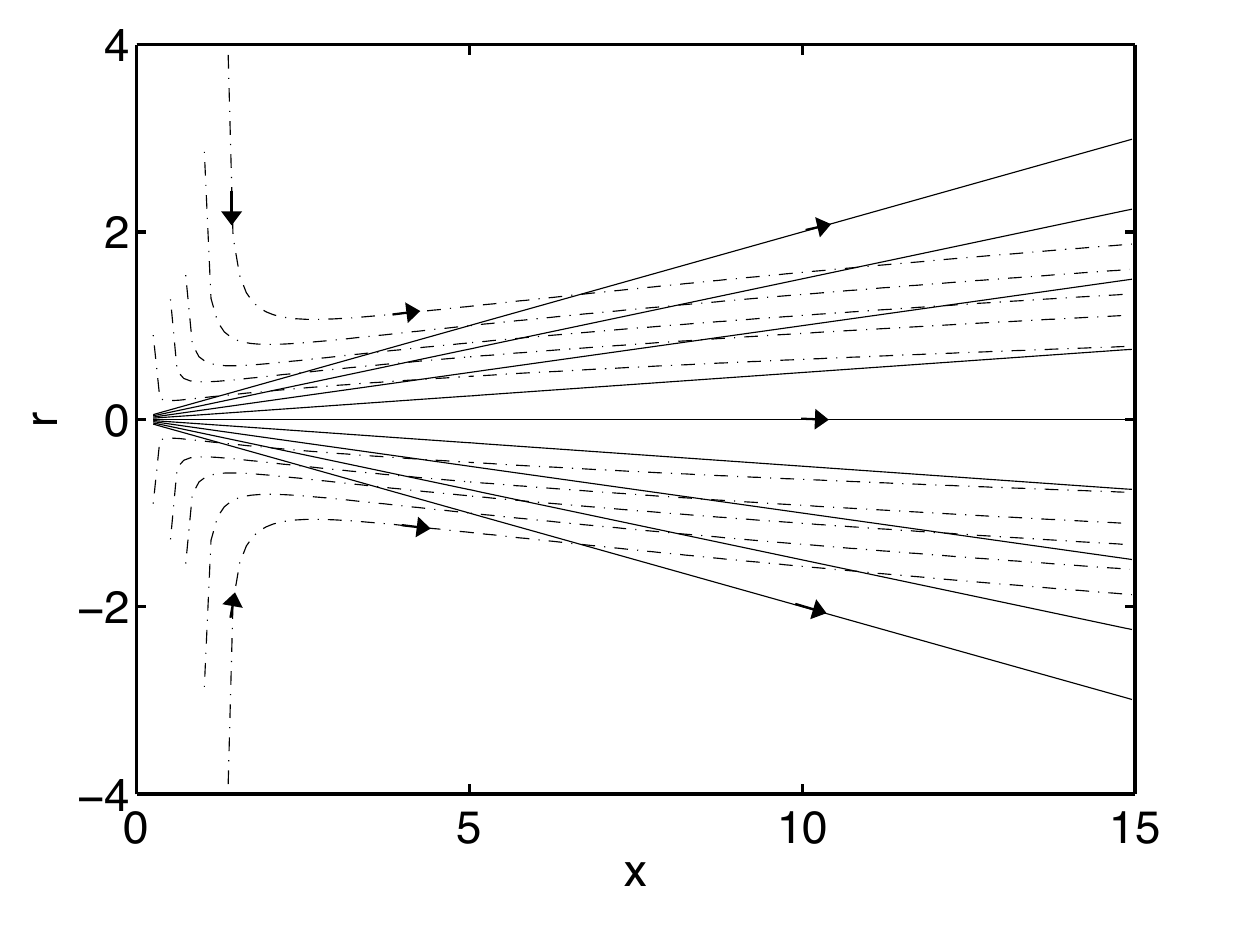}
  \end{center}
  \caption{Representative streamlines (dot-dashed) and momentum lines (solid) for laminar round jet similarity solution.}\label{f:roundjet-lines}
\end{figure}

The similarity solution for laminar round jet is given by the stream-function $\psi=\nu x f(\eta)$, where $\eta=r/x$ \citep{White2006}. The axial velocity is $u=(\nu/r) f^\prime$ and the radial velocity is $v=(\nu/r) (\eta f^\prime-f)$. The classical solution is $f(\eta)={(c\eta)^2}\left[{1+(c\eta/2)^2}\right]^{-1}$ where $c$ is related to the jet momentum flux and Reynolds number. The shape of the constant-$\psi$ streamlines are visualized in Fig. \ref{f:roundjet-lines} using dash-dotted lines. The horizontal momentum lines can be obtained using $F_{m,x}=u^2$ and  $F_{m,r} = uv - \nu \Pdt[u]{r} = (\nu/r)^2 f^{\prime} (\eta f^\prime - f +1- \eta f^\prime/f^{\prime\prime})$.
The slope of momentum lines is then given by
\begin{equation}
\frac{dr_m}{dx} =  \frac{F_{m,r}}{F_{m,x}} =  \eta + \frac{f^\prime-f f^\prime-\eta f^{\prime\prime}}{{f^\prime}^2}   = \eta,
\end{equation}
since $f^\prime-f f^\prime-\eta f^{\prime\prime}=0$ for the round-jet similarity solution. As a result, the momentum lines are straight lines (solid lines in Fig. \ref{f:roundjet-lines}). This is of course expected, since  momentum flux is constant in sections of cones bounded by a fixed $\eta$, a basic requirement since $\int_0^{\eta} {f^\prime}^2 d\eta'$ only depends upon $\eta$. The lines in Fig.  \ref{f:roundjet-lines} are helpful in visualizing how momentum is being brought towards the outer entrained fluid.

The kinetic energy lines in the far field (using the boundary-layer approximation that $K=\frac{1}{2} u^2$ ($v<<u$) also are lines but have twice the slope of the momentum lines (for the same value of $\eta$). The ``faster'' spreading does not mean that comparatively more kinetic energy than momentum arrives at the entrained fluid, since part of the kinetic energy is dissipated.

The far-wake solution with horizontal velocity $u=U_0-u_c$ and centerline defect velocity of $u_{c}=(C/x)\exp(-\eta^2)$ with $\eta=r [U_0/(4\nu x)]^{1/2}$ leads to a radial inflow given by $v=(\nu/U_0)^{1/2} C \, x^{-3/2} \, \eta \exp(-\eta^2)$. The radial momentum flux  is $F_{m,r} = uv - \nu \Pdt[u]{r}  \approx {U_0} v + {\nu} \Pdt[u_c]{r}$ in the far-wake, which can be shown to lead to  $v_m = 2v$. Similarly,  the kinetic energy transport vertical velocity is given by $v_K = 3 v$. Hence, momentum and kinetic energy are being transported into the wake from the outside along steeper, but similar, influx ``trajectories''. In the wake, mass, linear momentum and kinetic energy are being replenished.

\section{Transport of mass, momentum, and energy, in large wind farms}\label{s:windfarm}
We now turn to the application of transport tubes for the visualization of momentum and energy transport in the flow through wind-farm boundary layers.

With the increase in size of land-based and offshore wind farms the problem of farm performance is becoming an important research topic \citep{emeis1993,Frandsen2009,Ivanell2009,Barthelmie2010,Cal2010,Lu2011,Meyers2012}. For very large systems, the notion of the asymptotically large (infinite) wind farm becomes relevant \citep{emeis1993,Frandsen2006,Calaf2010}. This limiting case can be conveniently studied in numerical simulations using periodic boundary conditions in the horizontal direction, as has been done in recent LES studies of wind farms (\citealp{Calaf2010,Meyers2012}).

For a lone-standing turbine, physical mechanisms related to power extraction are reasonably well described using a stationary stream-tube analysis, neglecting effects of viscosity, and Reynolds stresses. Conservation of mass, the Bernoulli equation, and considering differences in up-stream and down-stream momentum fluxes,  lead to concepts such as the Betz limit for wind-turbine power extraction, wind-turbine momentum theory, etc. \citep{burton2001}. In real wind farms, however, wind-farm induced turbulence levels are much higher, so that turbulent fluxes become already as important as ideal terms in stream tubes that extend $2D$ upstream and downstream of turbines \citep{Lebron2012}. In such situations, the energy extracted by the turbines is entrained mostly from  the flow above the farm by turbulence, as quantified by Reynolds stress-mediated fluxes \citep{Calaf2010,Cal2010}. In order to help improve our understanding of the three-dimensional structure of these fluxes, in this section we investigate transport tubes of mass, momentum, and energy for eight different wind-turbine-array boundary layers, with different turbine spacings and configurations.

\begin{figure}
 \begin{center}
  \includegraphics[width=1\textwidth]{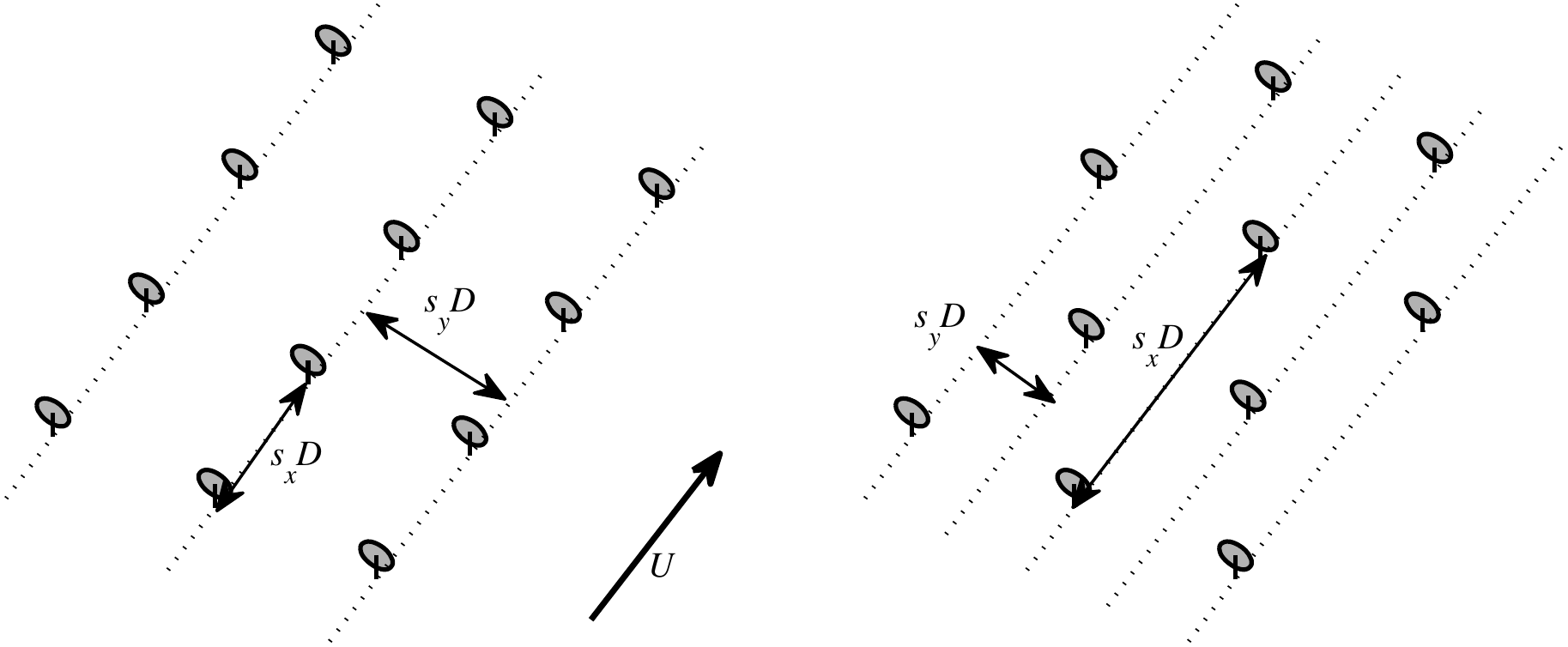}
  \end{center}
  \caption{Aligned (left) and staggered (right) turbine arrangement patterns, and definitions of stream-wise turbine-spacing $s_x$ and span-wise turbine spacing $s_y$ (non-dimensionalized by rotor diameter $D$). The mean flow ($U$) is in the $x$-direction}\label{f:Sketch}
\end{figure}

Table~\ref{tab:cases} provides an overview of the different cases considered. Four cases use an aligned arrangement pattern, while four other cases use a staggered pattern (see Figure~\ref{f:Sketch}  for a sketch). These cases comprise different stream-wise spacings $s_x D$ and span-wise spacings $s_y D$ between turbines (with $D$ the rotor diameter), as further detailed in Table 1. All results are obtained using LES following the approach discussed by \cite{Calaf2010,Meyers2010}. The four aligned cases are taken from \cite{Calaf2010}; the staggered cases are added in the current work, and are geometrically constructed by shifting every second span-wise row of turbines of the respective aligned cases along the span-wise direction. As a result, the stream-wise spacing between turbines doubles, while the span-wise spacing between rows is divided by two (see Figure~\ref{f:Sketch}).

\begin{figure}
  \begin{center}
  \includegraphics[width=0.6\textwidth]{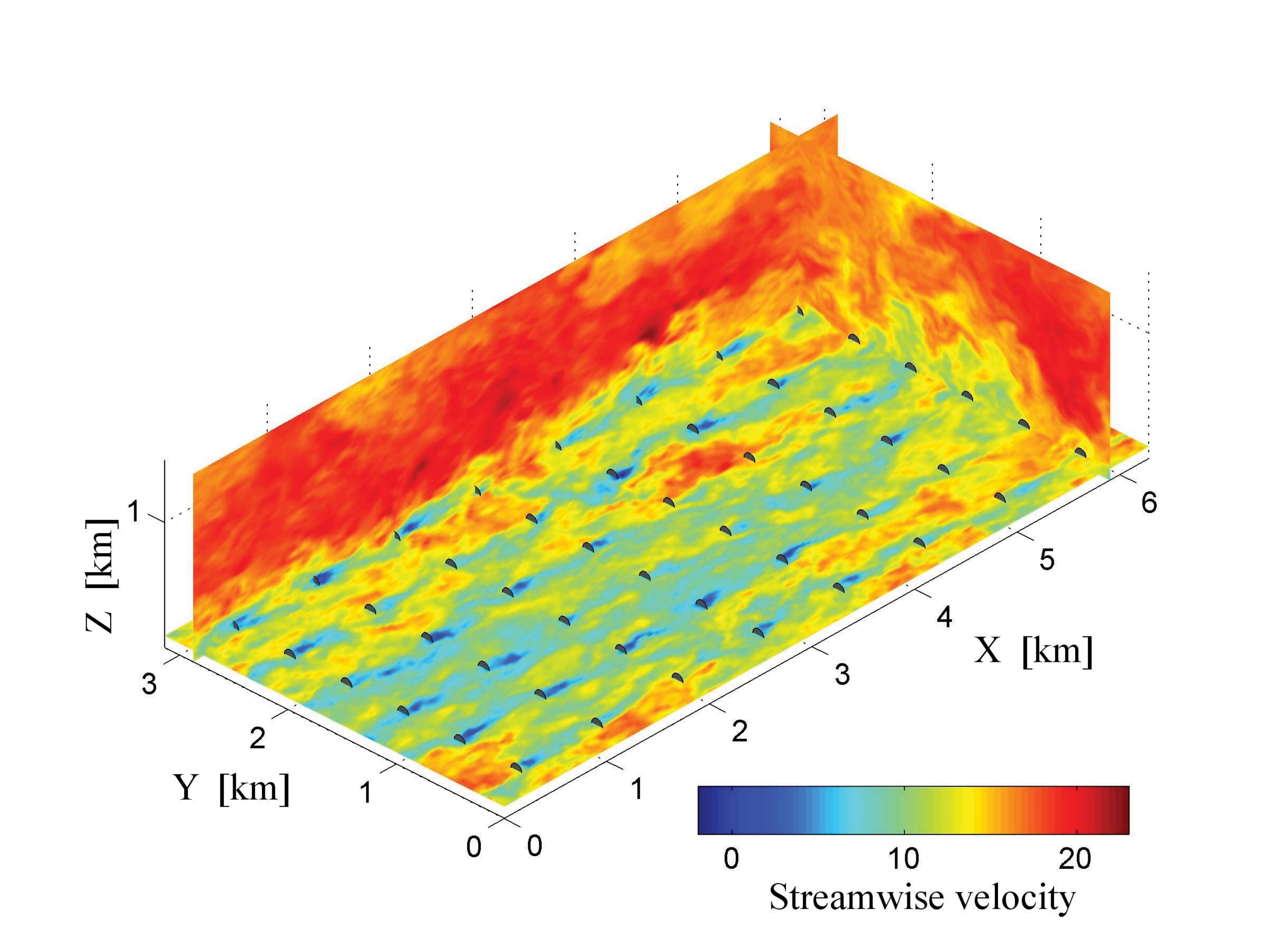}
  \end{center}
  \caption{Snapshot of a stream-wise velocity field in LES of a wind-turbine-array boundary layer. The color scale is stream-wise velocity in units of $u_* = [-(H/\rho)dp_{\infty}/dx]^{1/2}$.}\label{f:Snapshot}
\end{figure}

The effect of wind-turbines in the LES is represented using an Actuator Disk Model (ADM). We consider cases with and without wake rotation. Cases without wake rotation do not include applied tangential forces at the turbine (ADM, i.e Case 1--8 and 1F in Table~1). In another case (ADMR -- Case~1R in Table 1) we add tangential forces following the formulation used in \cite{Meyers2010}. In a recent detailed validation study by \cite{Wu2011}, it was demonstrated that except for near-wake effects close to the turbines with $x < 3D$, the non-rotating model (ADM) allows an accurate representation of the overall wake structures behind turbines. The rotating case (ADMR), including tangential forces, further improves near-wake behavior \citep{Wu2011}. Moreover, in the same study, the Reynolds stresses were found to be accurately predicted by both formulations, thus allowing an accurate representation of the interaction of the wind farms with the atmospheric boundary layer. A snapshot of a typical LES velocity field using the ADM method is provided in Figure~\ref{f:Snapshot}. Further details on the methodology and computational set-up may be found in \cite{Calaf2010} and \cite{Meyers2010}, and are summarized in Appendix~\ref{s:windfarmsimulations}, where the effects of LES resolution (Case~1 versus Case~1F), and wake rotation (Case~1 versus Case~1R) are also documented and discussed.

\begin{table}
\begin{center}
\caption{Turbine arrangement parameters, and spacing of the cases considered for evaluating transport tubes in the current work (see Figure~\ref{f:Sketch} for definition of stream-wise and span-wise turbine spacings). Number of turbines $N_t$ in the simulation domain, domain size $L_x \times L_y \times H$, and resolution $N_x\times N_y \times N_z$ of the LES are also shown (further details on the set-up, etc., are found in \citealp{Calaf2010}, and in Appendix B).}
\label{tab:cases}
\renewcommand{\arraystretch}{1.6}
\begin{tabular}{l@{\hspace{2mm}}l@{\hspace{2mm}}|@{\hspace{2mm}}l@{\hspace{4mm}}l@{\hspace{4mm}}l@{\hspace{4mm}}l@{\hspace{4mm}}l@{\hspace{4mm}}l}
  &  & $s_x$  & $s_y$  & $N_t$ & $L_x\!\times\! L_y\!\times\! H$ &$N_x\!\times\! N_y\!\times\! N_z$ & turbine model  \\
 Case 1 & aligned  & 7.85   & 5.24  & $8\!\times\! 6$ & $2\pi \!\times\! \pi \!\times\! 1$ & $128 \!\times\! 192 \!\times\! 61$  & ADM \\
 Case 2 & aligned  & 6.41  & 6.41 & $10\!\times\! 5$ & $2.04\pi \!\times\! 1.02\pi \!\times\! 1$& $128\!\times\! 192 \!\times\! 60$ & ADM\\
 Case 3 & aligned  & 9.07  & 4.54  & $7\!\times\! 7$ & $2.02\pi \!\times\! 1.01\pi \!\times\! 1$& $128\!\times\! 192 \!\times\! 61$ & ADM  \\
 Case 4 & aligned  & 15.7  & 10.5  & $4\!\times\! 3$ & $2\pi \!\times\! \pi \!\times\! 1$ & $128\!\times\! 192 \!\times\! 61$ & ADM\\
 \hline
 Case 5 & staggered & 15.7   & 2.62  & $8\!\times\! 6$ & $2\pi \!\times\! \pi \!\times\! 1$ & $128 \!\times\! 192 \!\times\! 61$   & ADM\\
 Case 6 & staggered & 12.8   & 3.21 & $10\!\times\! 5$ & $2.04\pi \!\times\! 1.02\pi \!\times\! 1$& $128\!\times\! 192 \!\times\! 60$ & ADM\\
 Case 7 & staggered & 18.1   & 2.27  & $8\!\times\! 7$ & $2.31\pi \!\times\! 1.01\pi \!\times\! 1$& $128\!\times\! 192 \!\times\! 61$  & ADM \\
 Case 8 & staggered & 31.4   & 5.24  & $4\!\times\! 3$ & $2\pi \!\times\! \pi \!\times\! 1$ & $128\!\times\! 192 \!\times\! 61$ & ADM\\ \hline
 Case 1F & aligned   & 7.85   & 5.24  & $8\!\times\! 6$ & $2\pi \!\times\! \pi \!\times\! 1$ & $192 \!\times\! 320 \!\times\! 102$   & ADM\\
 Case 1R & aligned   & 7.85   & 5.24  & $8\!\times\! 6$ & $2\pi \!\times\! \pi \!\times\! 1$ & $128 \!\times\! 192 \!\times\! 61$ & ADMR
 \end{tabular}
 \renewcommand{\arraystretch}{1.0}
\end{center}
\end{table}

\begin{figure*}
\begin{center}
   \includegraphics[width=1\textwidth]{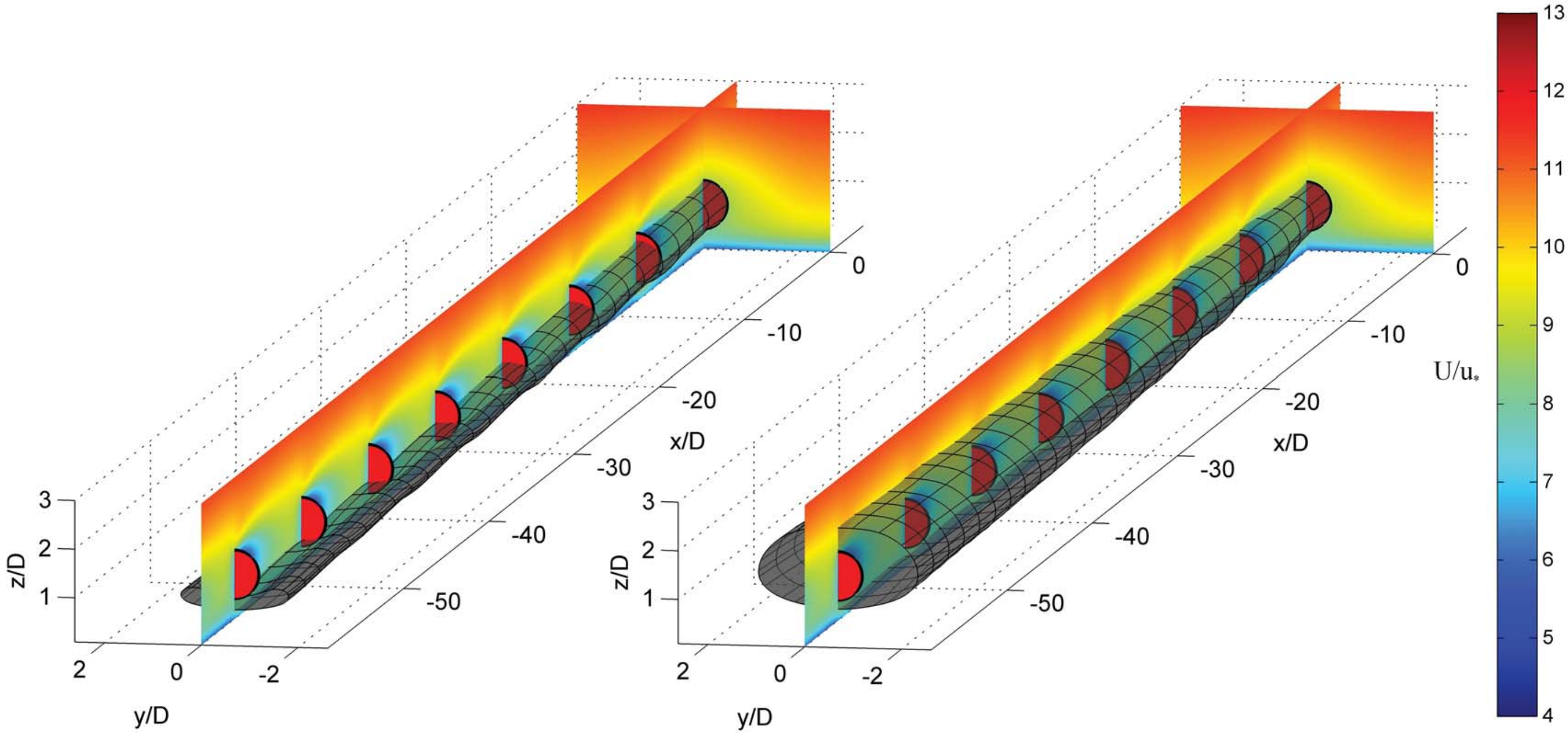}
 \end{center}
\caption{Mean stream-wise velocity field, (left) stream tube, and (right) total mechanical energy tube in a turbine row of a fully developed wind-turbine-array boundary layer flow (corresponding to Case 1 in Table~\ref{tab:cases}). The color scale is stream-wise velocity in units of $u_* = [-(H/\rho)dp_{\infty}/dx]^{1/2}$.}\label{f:visualizetubes}
\end{figure*}

\begin{figure*}
  \begin{center}
  \includegraphics[scale=0.48]{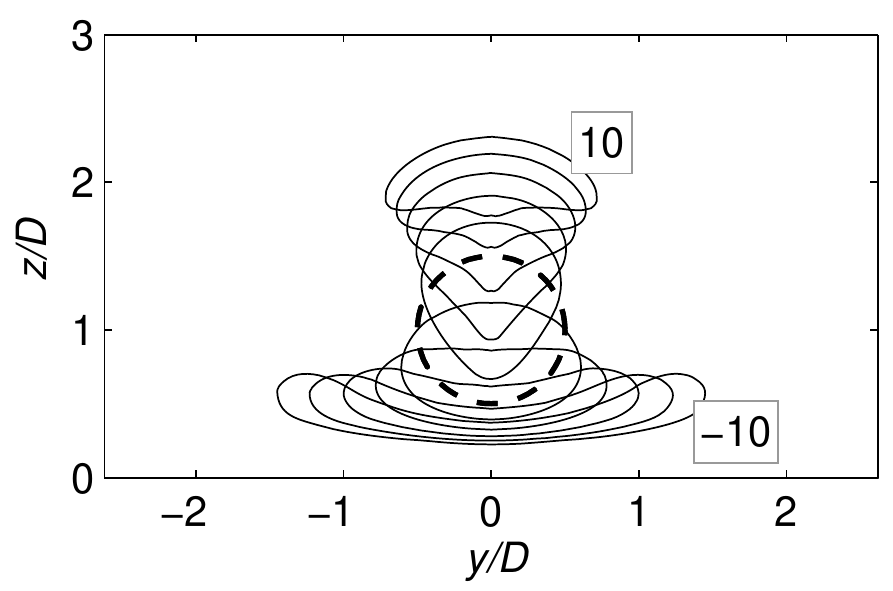}{\scriptsize (a)}
  \includegraphics[scale=0.48]{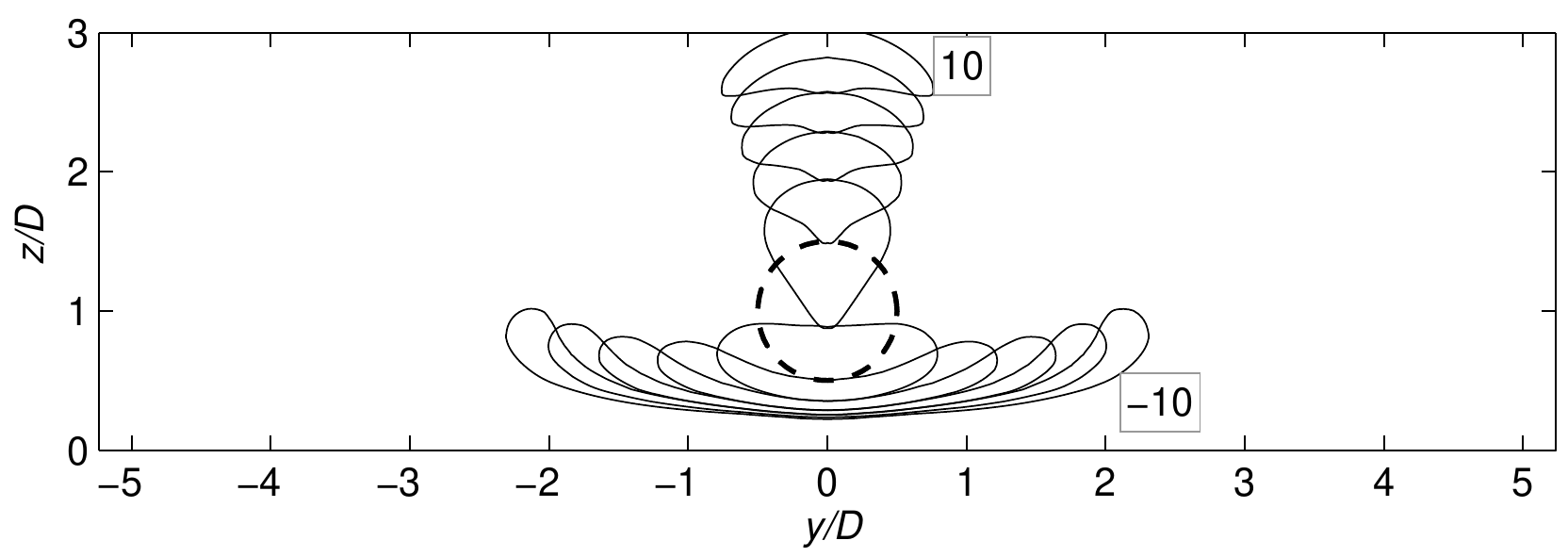}{\scriptsize (b)}
  \includegraphics[scale=0.48]{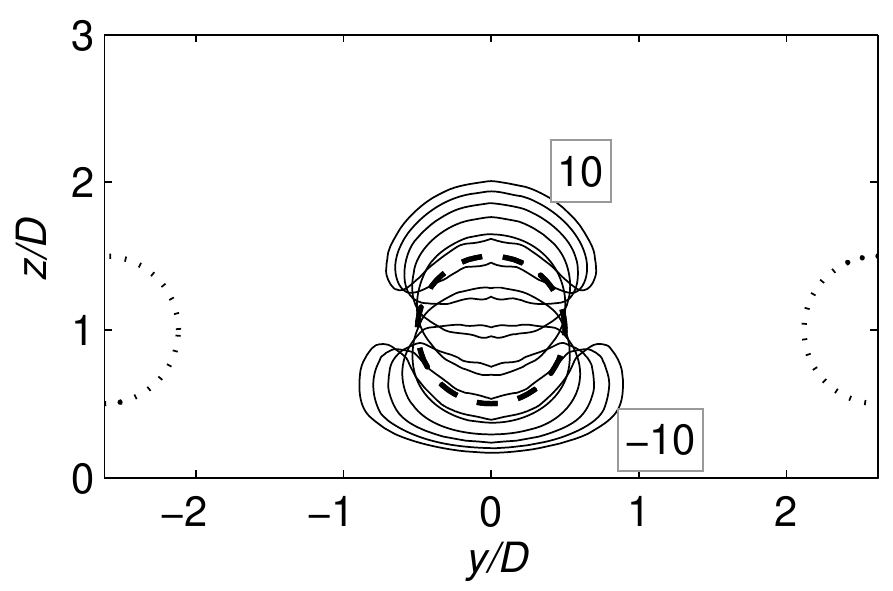}{\scriptsize (c)}
  \includegraphics[scale=0.48]{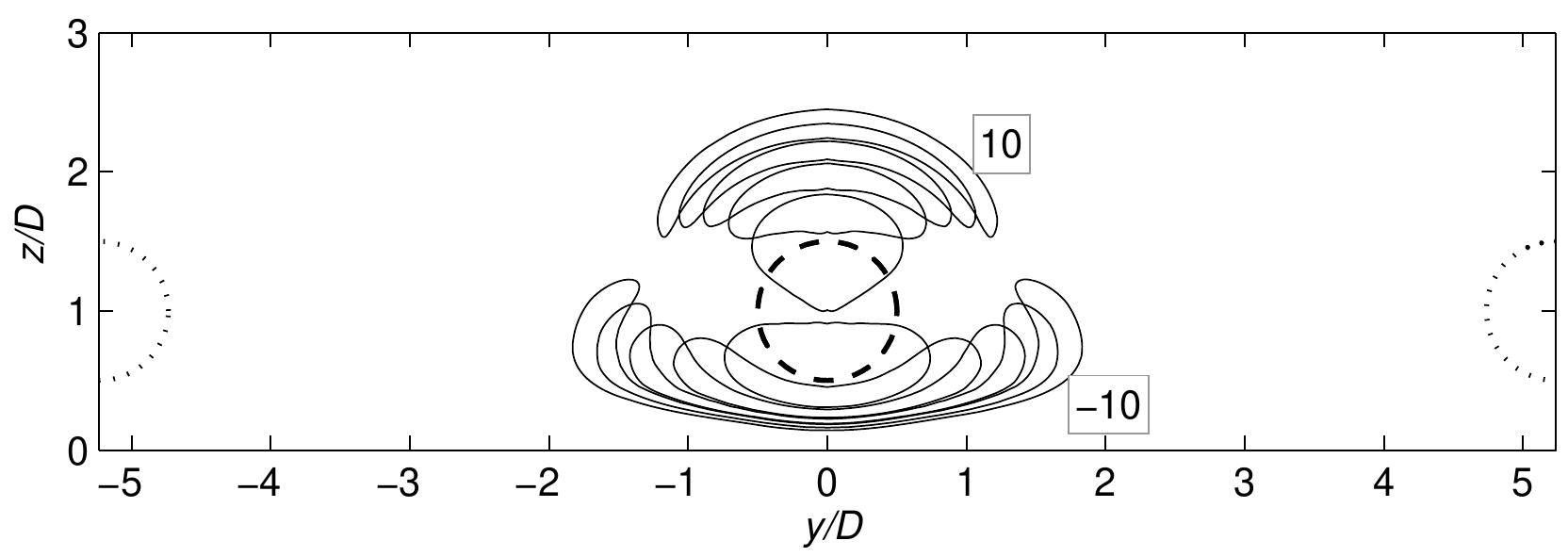}{\scriptsize (d)}
  \end{center}
\caption{Upstream and downstream sections of stream tubes for wind farms with different turbine spacings. (a-b) aligned Cases 1, and 4; and (c--d) staggered Cases 5 and 8 -- see Table~\ref{tab:cases} for details. ($--$): turbine rotor; (---): sections at different upstream and downstream locations, with distances corresponding to $x=\pm n s_x D$, and $n=2,4,\cdots, 10$ (farthest sections, at $n=\pm 10$, are labeled). ($\cdots$): in (c)--(d) corresponds to the location of the staggered row of turbines.}\label{f:streamtube}
\end{figure*}

\subsection{Results: mean-flow stream tubes}\label{s:masstubes}
As a first step, we visualize classical stream tubes for Case 1. To this end, the three-dimensional velocity field is averaged in time to obtain a spatially periodic mean-flow velocity field with period corresponding to the turbine spacing, i.e. $7.85D$ in stream-wise, and $5.23D$ in span-wise directions for Case 1. Stream-tubes are obtained by constructing streamlines through 60 equally spaced seed points along a circle that coincides with the target turbine disk. The streamlines are tangent to the mean-velocity vector field obtained from LES (or later to the vector fields given by \ref{eq:defomflux}, or \ref{eq:defue}). During the procedure, we regularly add seed points whenever the curvature of downstream or upstream cross sections becomes too large, or stream lines are too widely spaced along the tube mantle.

In Figure~\ref{f:visualizetubes}(a) we show the mean stream-wise velocity field in a 8 by 1 turbine row, together with the turbine-rotor stream tube through the downstream turbine, i.e., defined by the streamlines through the rotor disk of that turbine. Note that the velocity field is periodic, but the stream tube is not.  In Figure~\ref{f:visualizetubes}(b) the energy tube is also shown for Case~1, illustrating large differences between both types of tubes. The differences are due to the considerable transport across the stream tube associated with turbulence (Reynolds stresses). \cite{Lebron2012} measured such fluxes through the mantle of a stream-tube using wind-tunnel data from a model wind farm, and found that the turbulent fluxes were dominant.  Further discussion of this case is continued below, but first, we present stream tubes for the other wind-farm cases introduced in Table~\ref{tab:cases}. To this end, we display sections of rotor-disk stream tubes at upstream and downstream rotor planes in Figure~\ref{f:streamtube}(a)--(d) for a selected number of wind-farm cases. We observe, again, that the stream tube continuously deforms further and further away from the rotor disk, and that the tube center does not remain at hub height. It is appreciated from these periodic cuts (``Poincar\'{e} sections'') that the average mass flux through the turbine rotor plane is originating upstream from below the turbine level, while downstream it is ejected above turbine level. The main difference between the aligned cases (Figure~\ref{f:streamtube}(a),(b)) and the staggered cases (\ref{f:streamtube}(c),(d)), is that the stream tubes in the latter cases extend much less to the sides, as the sideways development of these tubes in constrained by the neighboring out-of-plane turbine rows. For sake of brevity, stream tubes of other cases are not shown here, as they have features which are very similar to the cases shown in Figure~\ref{f:streamtube}

\begin{figure*}
  \begin{center}
  \includegraphics[width=0.45\textwidth]{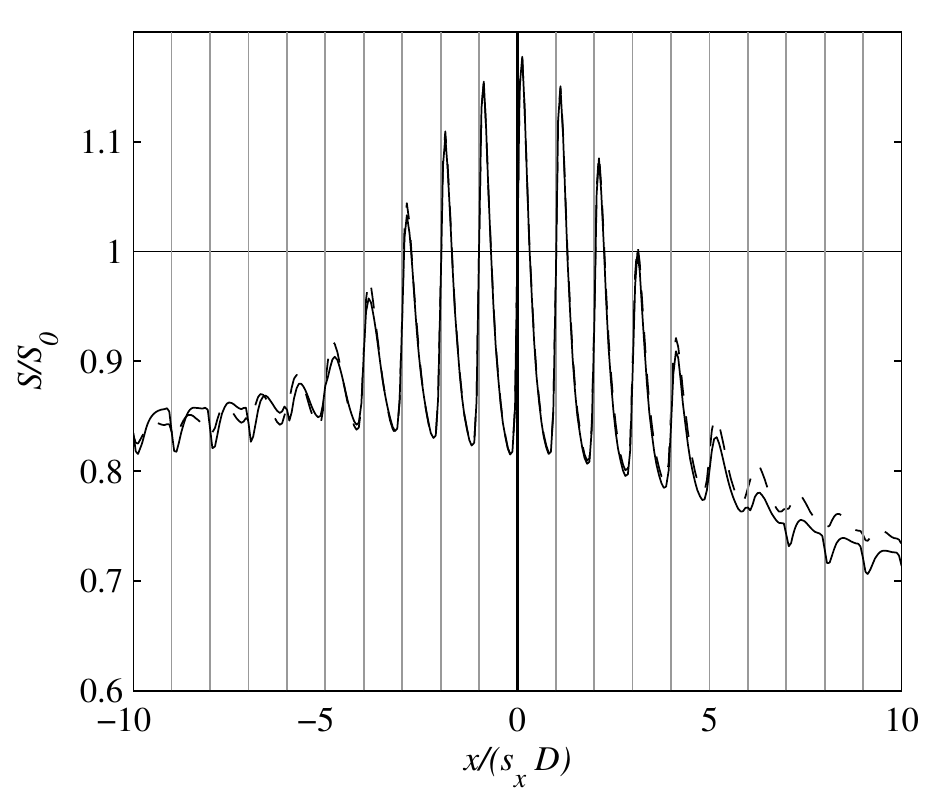}{\scriptsize(a)}
  \includegraphics[width=0.45\textwidth]{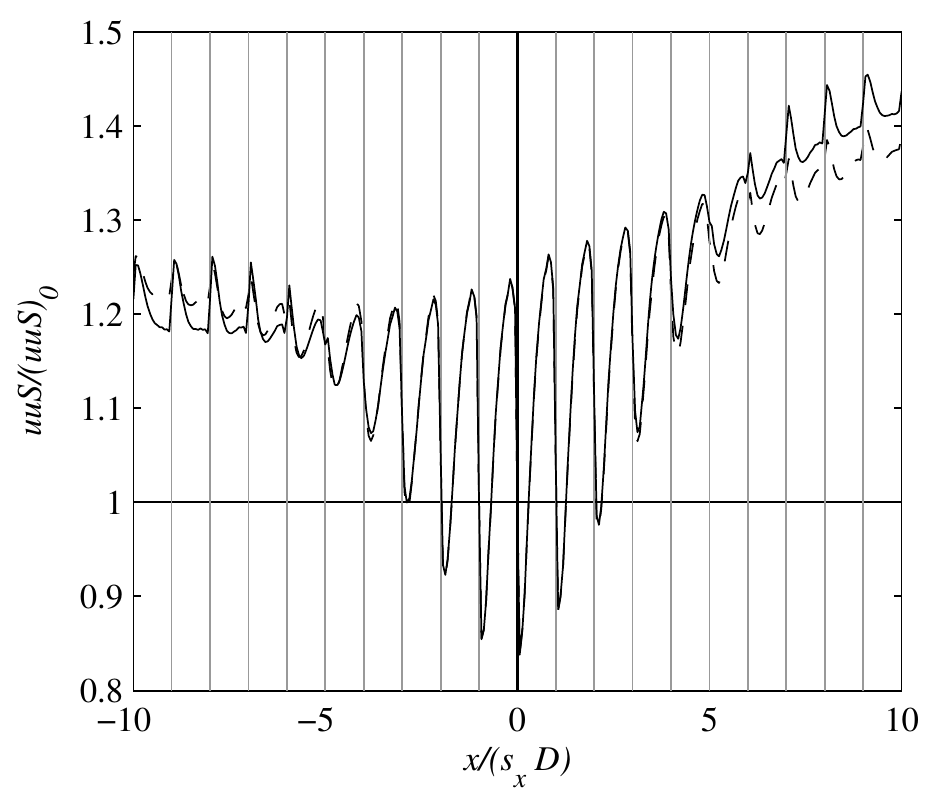}{\scriptsize (b)}
  \includegraphics[width=0.45\textwidth]{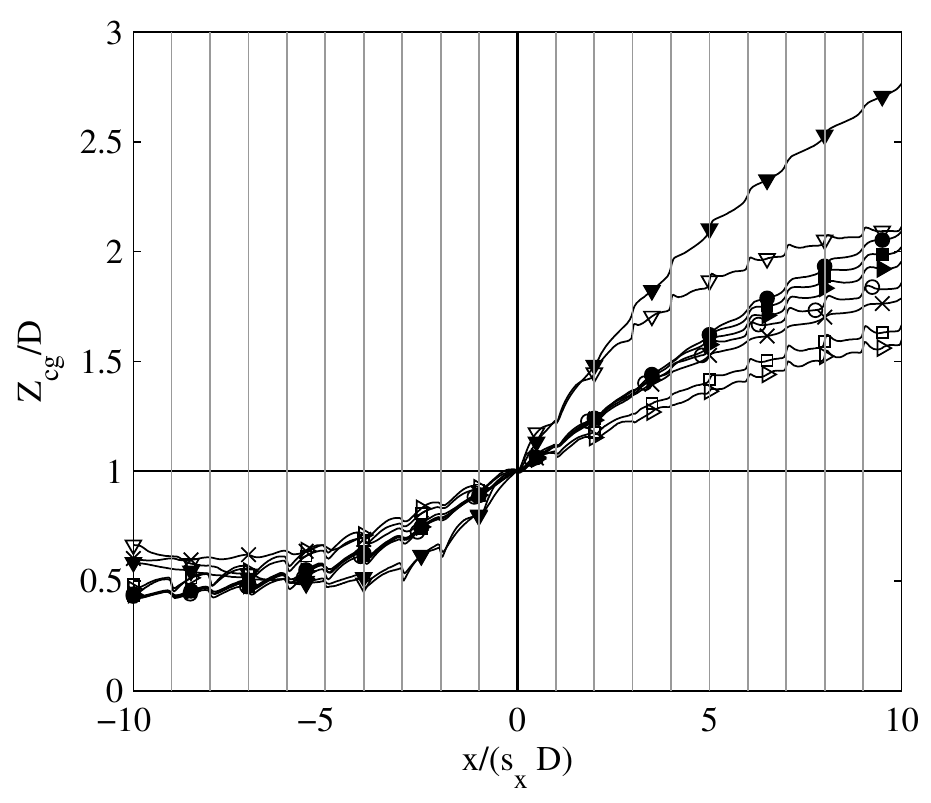}{\scriptsize (c)}
  \end{center}
\caption{(a,b) Evaluation of stream tubes in a wind-farm boundary layer with turbine spacing $s_x = 7.85$, and $s_y=5.23$ with (---): Case 1, and ($--$): Case 1R (see Table 1). (a) Surface area along a turbine-rotor stream tube. (b) Flux of mean axial momentum through the stream tube. (c) Geometrical center of gravity of the stream tubes (all cases, see Table 1), with closed symbols: aligned case; and open symbols: staggered case.  $\blacksquare, \square$: Case 1,  5; $\circ, \bullet$: Case 2, 6; $\blacktriangleright,\vartriangleright$: Case 3, 7; $\blacktriangledown,\triangledown$: Case 4, 8; and $\times$: Case 1R.  }\label{f:streamtubeeval}
\end{figure*}

Before turning to the main topic of momentum and energy transport tubes, first we report further properties of the classical stream tubes by evaluating the area ($S$, inversely proportional to the section-averaged mean velocity) of vertical cross sections of the tubes, as well as the axial fluxes crossing these sections as function of downstream distance. The evolution of $S$ and the axial momentum flux are plotted for Case 1 and Case 1R in Figure~\ref{f:streamtubeeval}(a) and (b). It is observed that the evolution of cross-sectional surface and momentum flux along the tubes is quite similar for both cases, with and without rotation. Overall, we find that effects of wake rotation do not dominate transport of momentum or energy, and further discussion, comparing Case 1 and Case 1R, is provided in Appendix~\ref{s:windfarmsimulations}.

In Figure~\ref{f:streamtubeeval}(a) the surface $S$ is displayed as function of the upstream and downstream distance from the tube's originating turbine disk. At $x=0$, it is observed that the slope of $S$ is positive, associated with a reduction of the flow velocity by the turbine disk thrust forces. Further downstream $0<x/(s_xD)<1$, the surface decreases again, related to a speed-up of the flow (wake recovery). Also in upstream and downstream turbine planes ($-6<x/(s_xD)<6$), similar trends are observed: at the turbine planes the area $S$ increases (slow down of the flow), in between turbines it decreases. It is further observed that the difference between maximum and minimum $S$ decreases farther from $x=0$, as the intersection of the stream tube with the turbine rotors at upstream and downstream planes decreases. Sufficiently far upstream (or downstream), e.g., $x/(s_xD) \leq -6$, the trends change. Now the tube area $S$ shrinks at turbine planes (with speed up of the flow), and grows in between (slow down). Here, the tube is not intersecting anymore with the turbine rotor. Hence, the average flow speeds up at the rotor, i.e. flow is partially driven around the rotor, and the flow slows down in between rotor planes, i.e. part of its momentum is transferred to the wake regions behind the turbine rotors by Reynolds-stress interactions.

In Figure~\ref{f:streamtubeeval}(b) the flux of axial momentum through the stream tube is shown. Trends observed can be explained using the same rationale as above, and are largely related to the effective intersection between the stream tube and the turbine rotor disk regions. We further observe that the maxima of fluxes of momentum through the tube increase for $-6<x/(s_xD)<10$. This is explained by the ascending trajectory of the stream tube, and the increase of mean-flow background momentum which is available to replenish momentum in the turbine wakes. Fluxes of energy through the stream tube, look very similar to the evolution of momentum fluxes, and are not further shown here.

From the analysis above and Figure~\ref{f:streamtubeeval}(a,b), it is appreciated that the flux of axial momentum through conventional stream tubes is highly non-trivial, affected by the upward motion of the tubes through the farm, together with Reynolds-stress exchanges over the tube mantle. Trends for the other cases (not shown) are the same. For all cases we find that through the mean velocity field, fluid volume (or mass) comes from below the turbines and downstream is ejected above the turbines. This is illustrated in Figure~\ref{f:streamtubeeval}(c), where the geometric center of the stream tube vertical cross sections is presented for all cases. We also observe that the geometric center for staggered cases remains closer to the turbine center. Note that the stream-wise turbine spacing $s_x$ in the staggered cases is twice that of the aligned cases, so that this difference between staggered and aligned effect would be even more pronounced when the stream-wise distance is not normalized by $s_xD$.

\subsection{Results: Momentum and energy tubes}
We now turn to the determination of momentum and energy transport tubes as defined in \S\ref{s:massmomenergy}. Since the molecular viscosity in the LES is set to zero, and the contribution of the sub-grid eddy-viscosity compared to the resolved stresses is negligible (except very close to the ground), we consider only the Reynolds stresses based on the resolved velocity field in defining the transport velocities.

\begin{figure*}
  \begin{center}
  \includegraphics[scale=0.5]{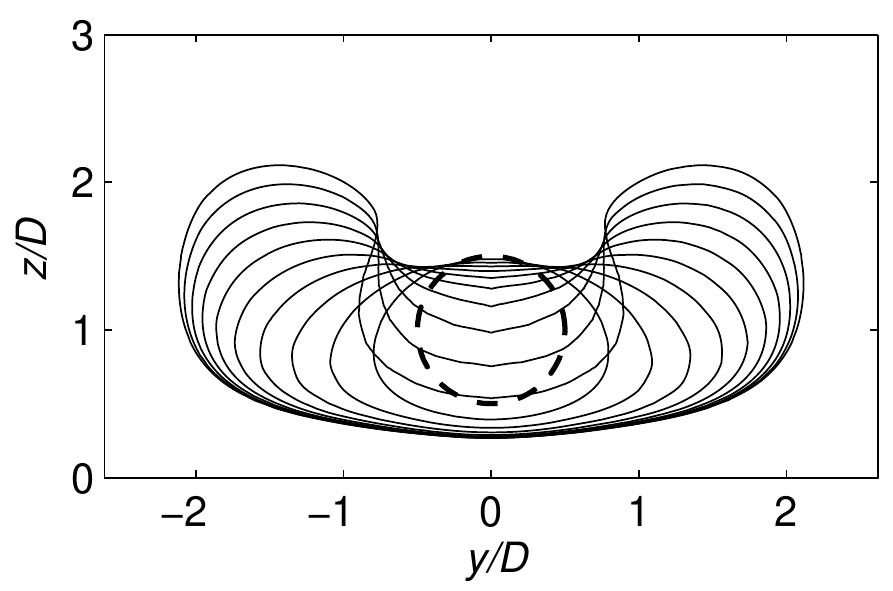}{\scriptsize (a)}
  \includegraphics[scale=0.5]{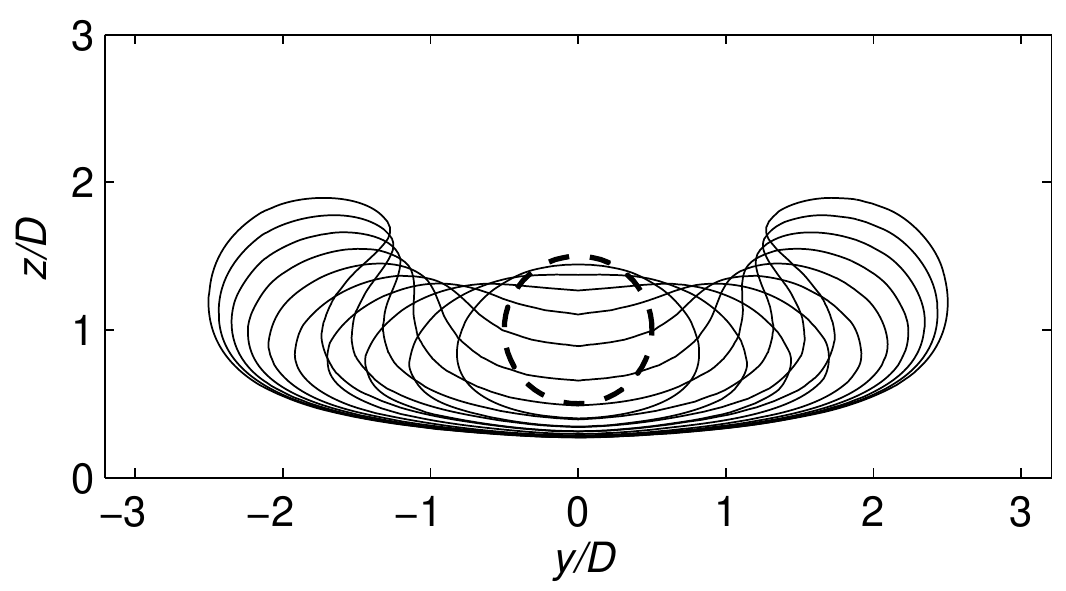}{\scriptsize (b)}
  \includegraphics[scale=0.5]{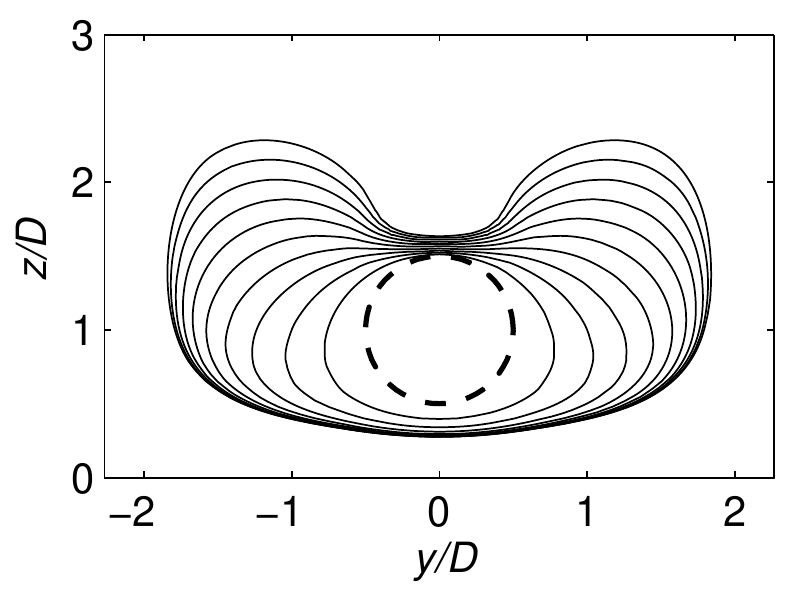}{\scriptsize (c)}
  \includegraphics[scale=0.5]{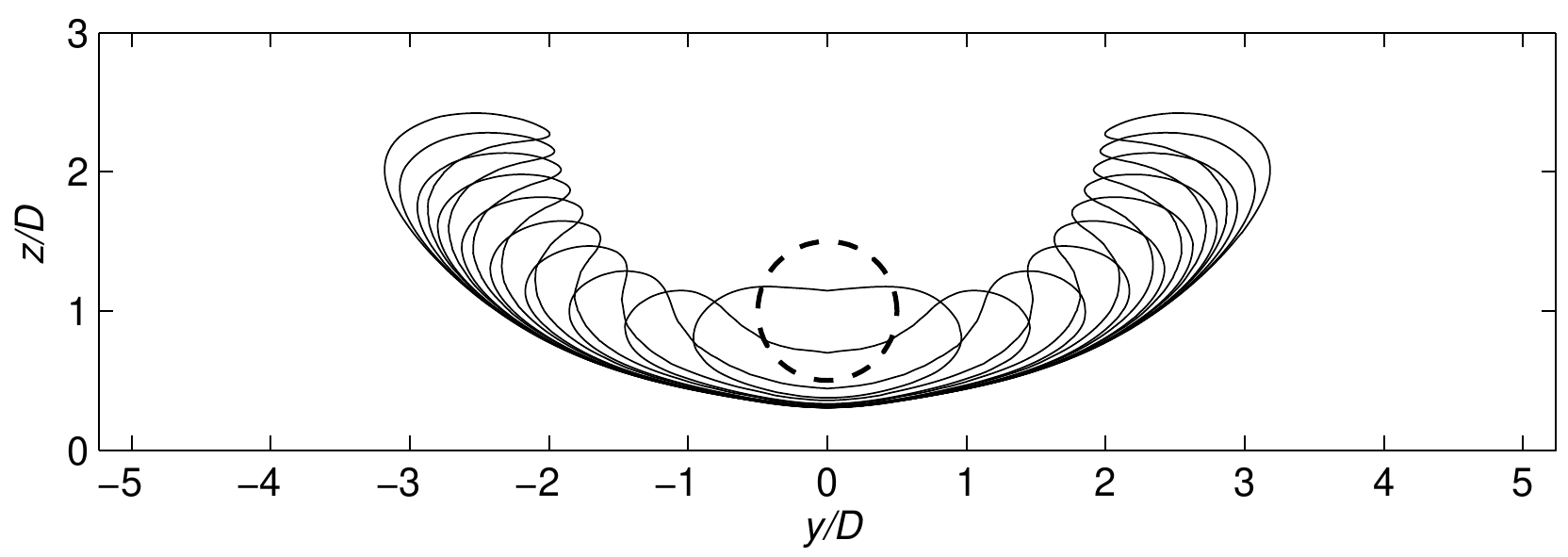}{\scriptsize (d)}
  \includegraphics[scale=0.5]{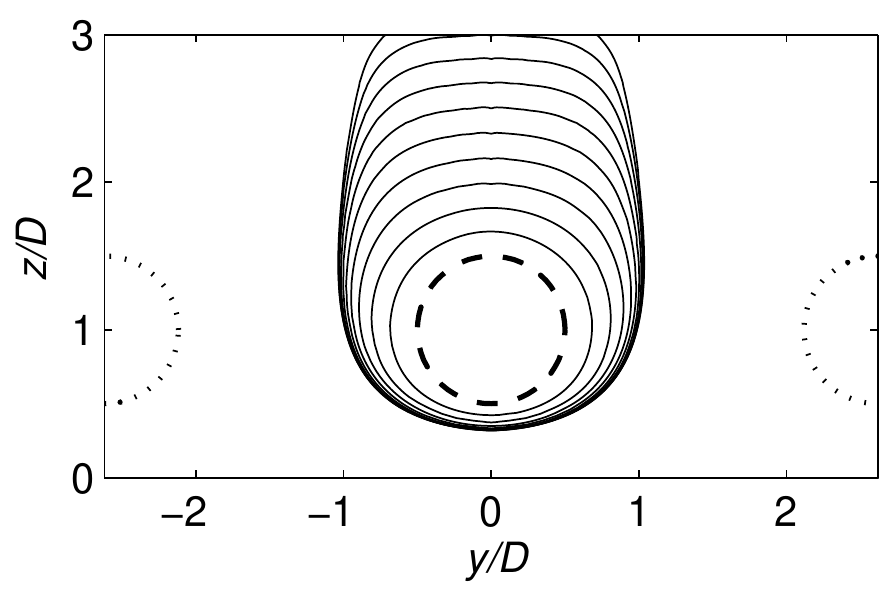}{\scriptsize (e)}
  \includegraphics[scale=0.5]{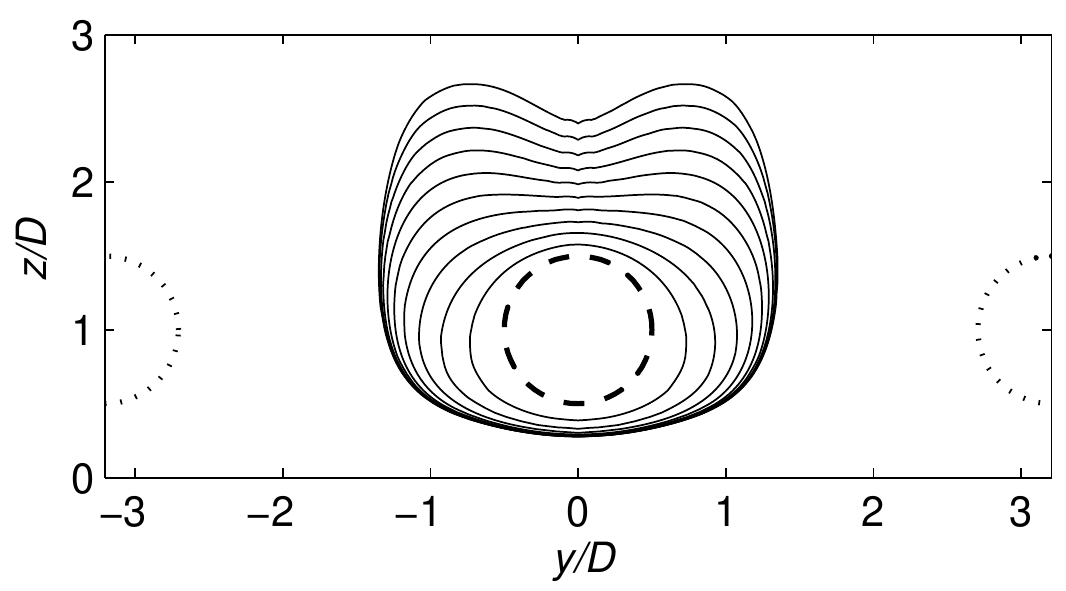}{\scriptsize (f)}
  \includegraphics[scale=0.5]{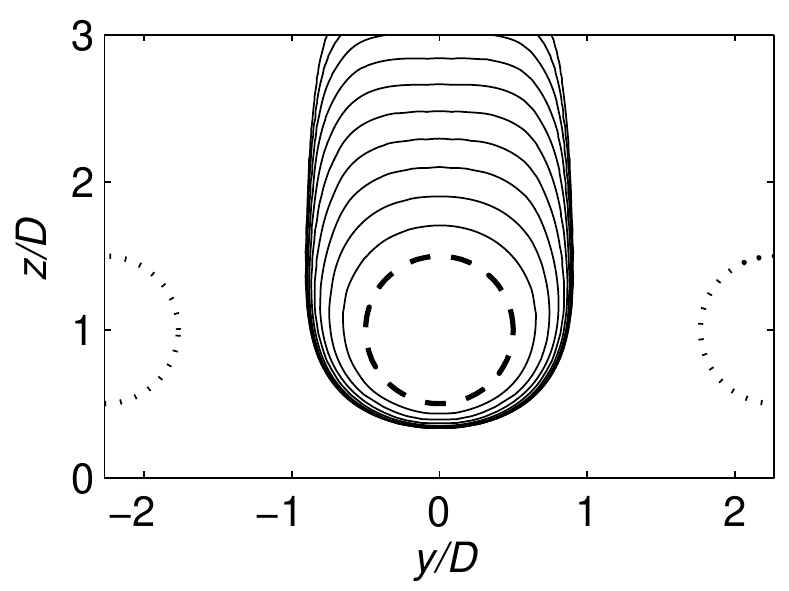}{\scriptsize (g)}
  \includegraphics[scale=0.5]{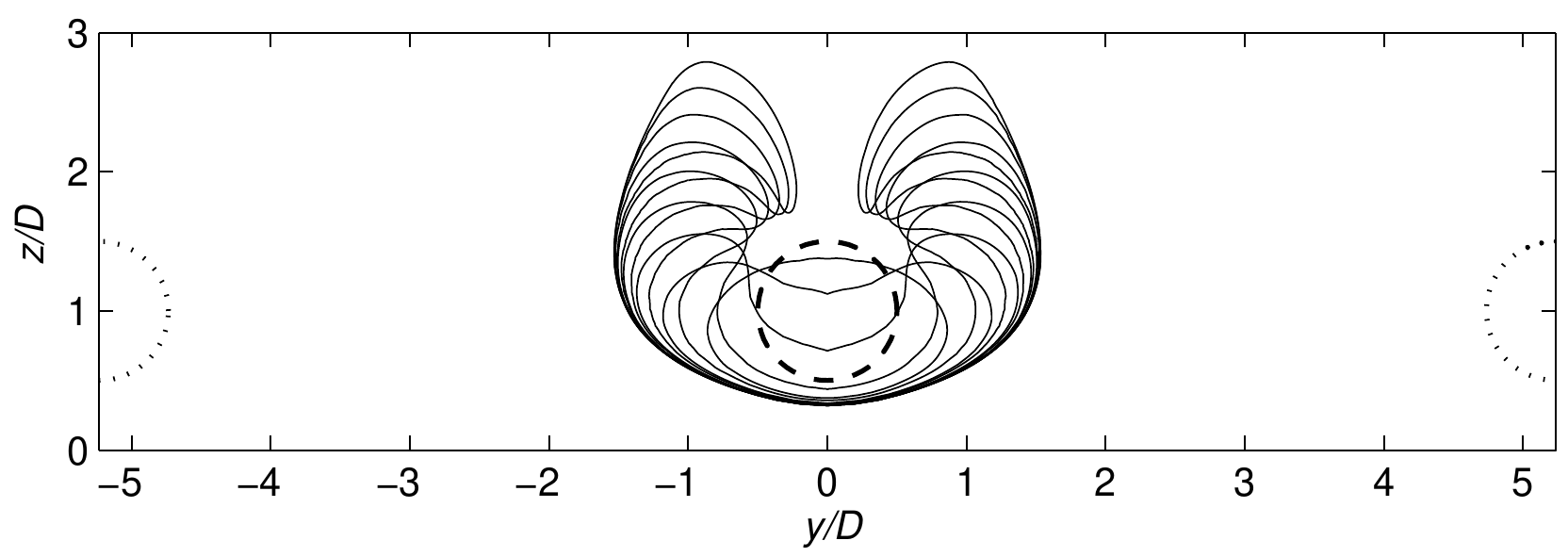}{\scriptsize (h)}
  \end{center}
\caption{Upstream sections of axial-momentum transport tubes for wind farms with different turbine spacings. (a--h) corresponds respectively with Case 1--8, where (a--d) are all aligned cases, and (e--h) are staggered cases -- see Table~\ref{tab:cases} for details. ($--$): turbine rotor; (---): sections at different upstream locations, with upstream distances corresponding to $x=-n s_x D$, and $n=2,4,\cdots, 20$. ($\cdots$): in (e)--(h) corresponds to the location of the staggered row of turbines.}\label{f:momentumtube}
\end{figure*}

\begin{figure*}
  \begin{center}
  \includegraphics[scale=0.5]{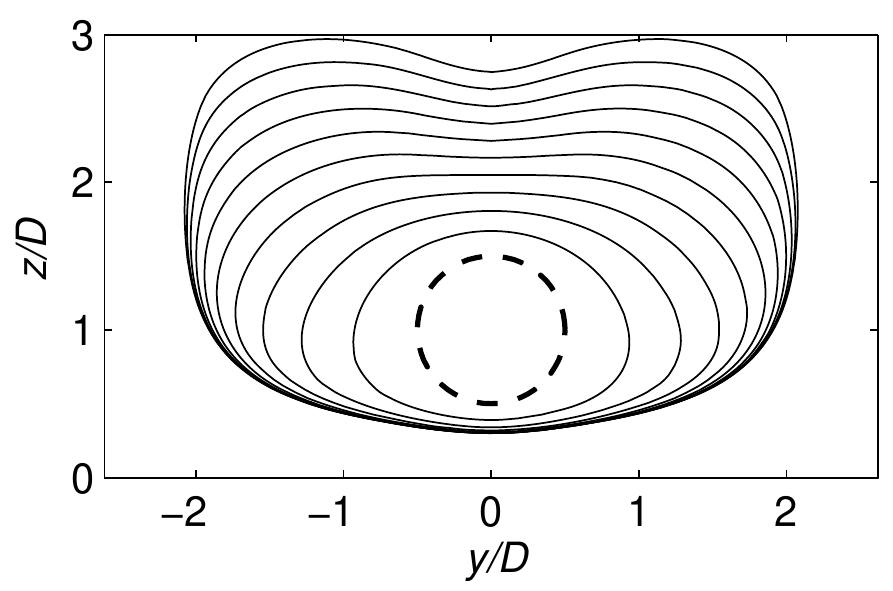}{\scriptsize (a)}
  \includegraphics[scale=0.5]{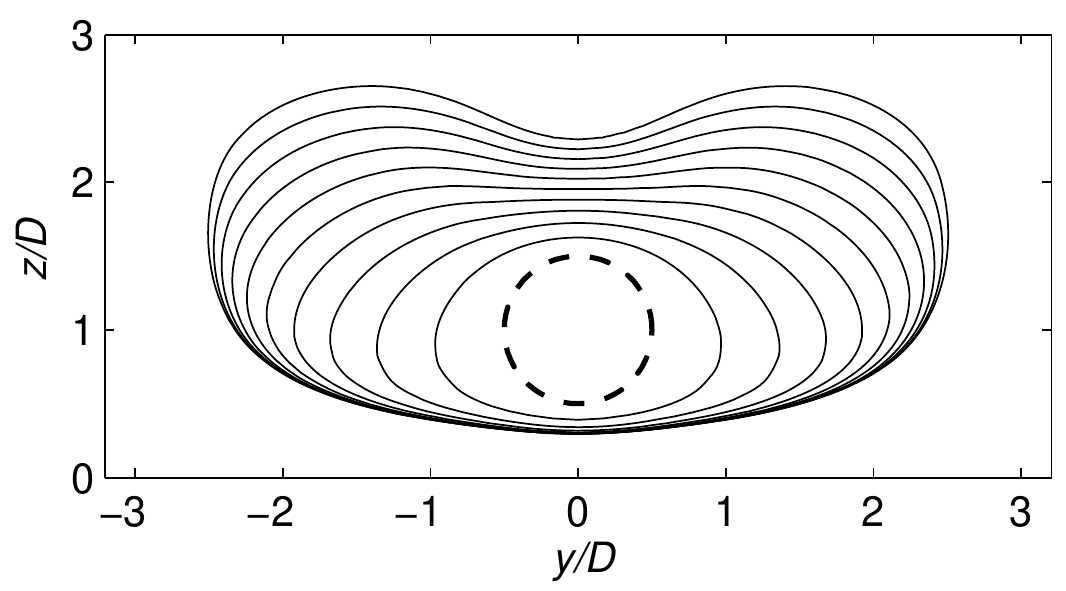}{\scriptsize (b)}
  \includegraphics[scale=0.5]{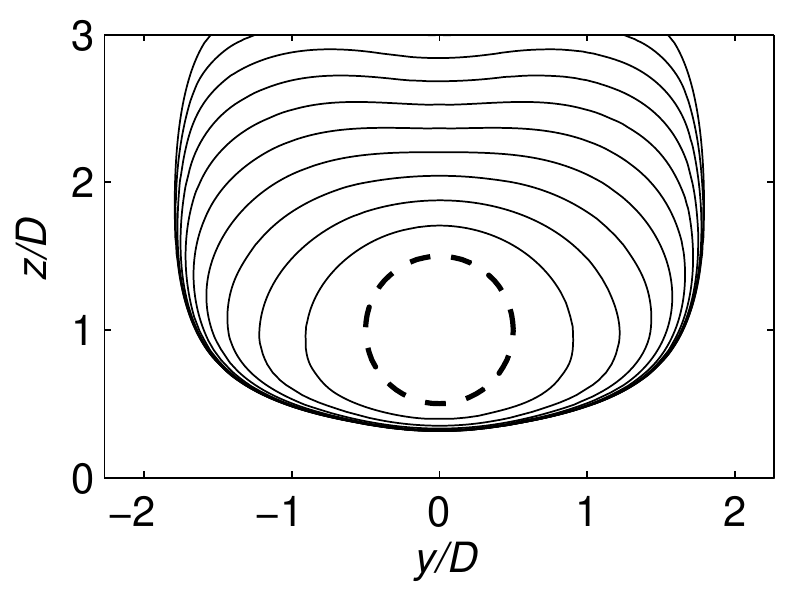}{\scriptsize (c)}
  \includegraphics[scale=0.5]{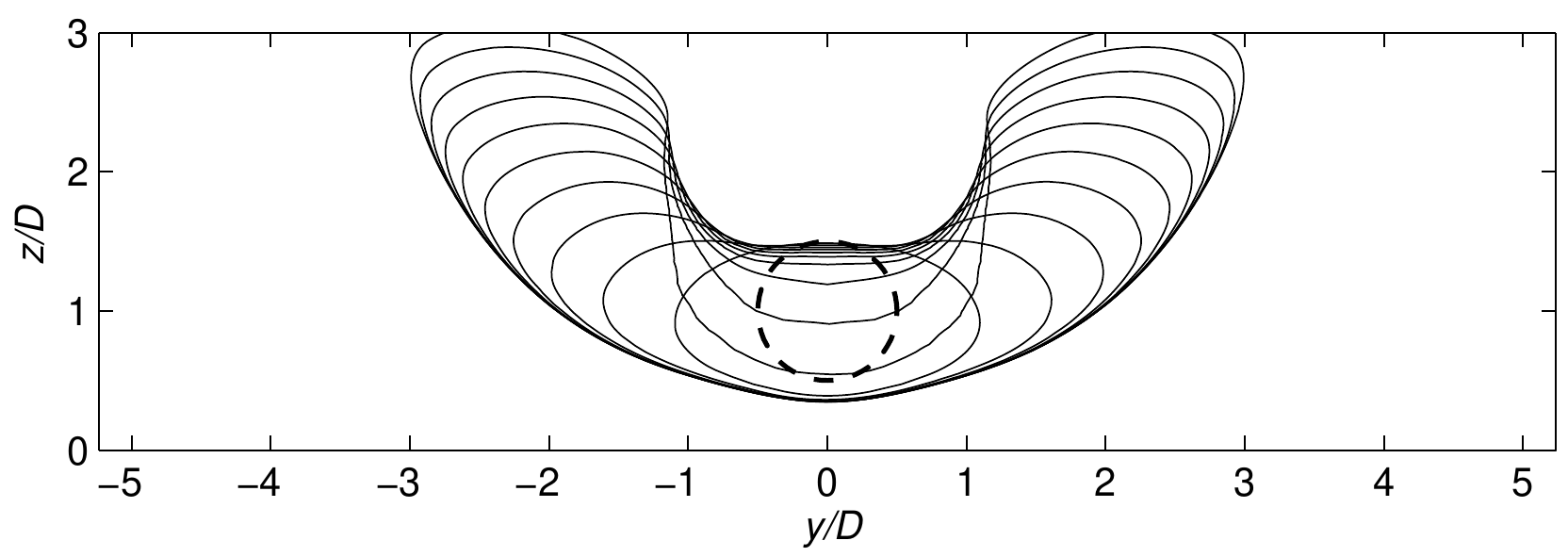}{\scriptsize (d)}
  \includegraphics[scale=0.5]{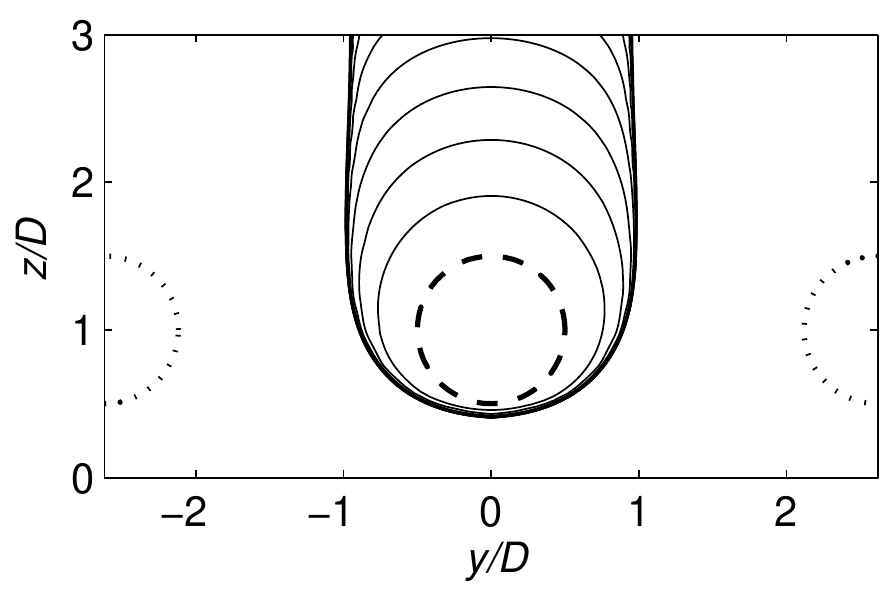}{\scriptsize (e)}
  \includegraphics[scale=0.5]{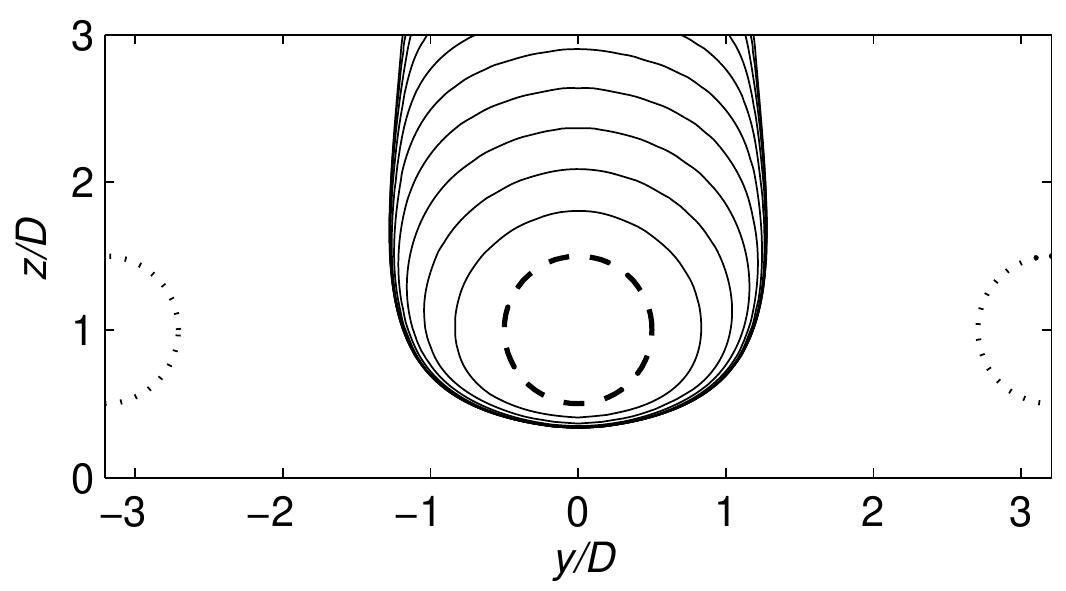}{\scriptsize (f)}
  \includegraphics[scale=0.5]{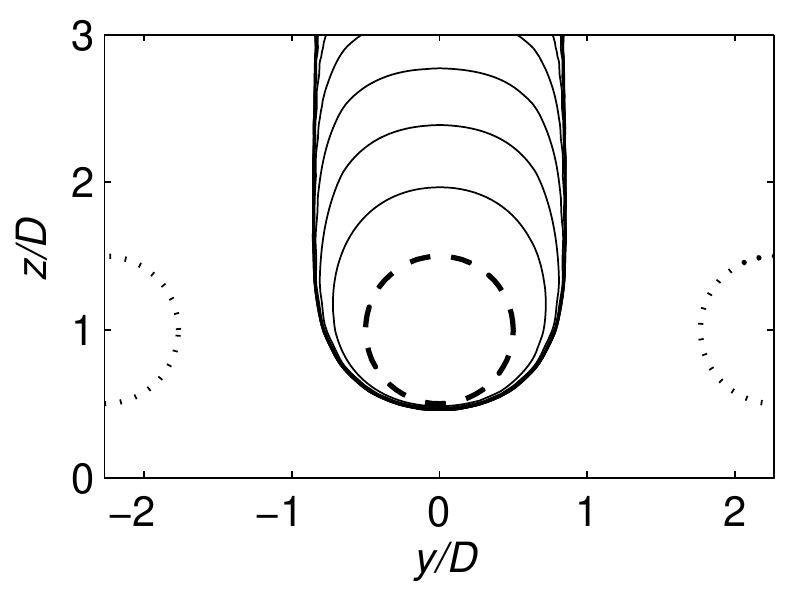}{\scriptsize (g)}
  \includegraphics[scale=0.5]{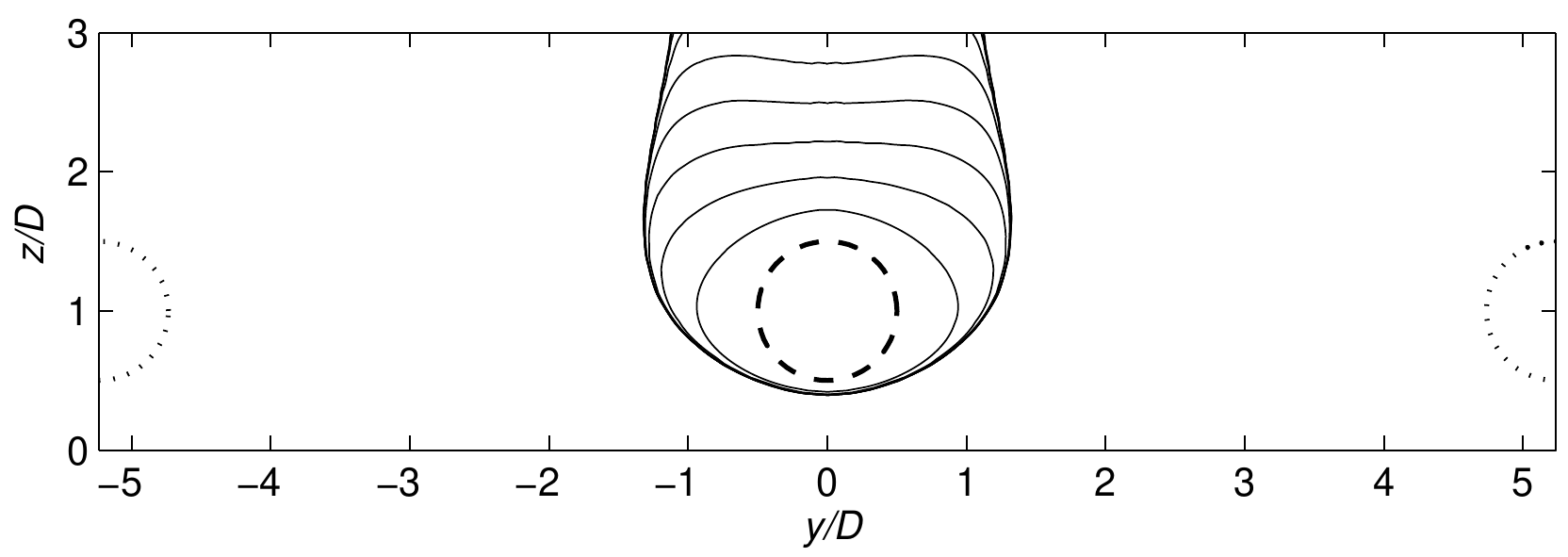}{\scriptsize (h)}
  \end{center}
\caption{Upstream sections of mean-flow mechanical energy transport tubes for wind farms with different turbine spacings. (a--h) corresponds respectively with Case 1--8, where (a--d) are all aligned cases, and (e--h) are staggered cases -- see Table~\ref{tab:cases} for details. ($--$): turbine rotor; (---): sections at different upstream locations, with upstream distances corresponding to $x=-n s_x D$, and $n=2,4,\cdots, 20$. ($\cdots$): in (e)--(h) corresponds to the location of the staggered row of turbines.}\label{f:energytube}
\end{figure*}

In Figures~\ref{f:momentumtube} and \ref{f:energytube} sections of transport tubes of axial momentum and mean-flow mechanical energy, respectively, are shown for the various cases of Table I.  These transport tubes are considerably different from the conventional stream tubes shown before. Neither momentum nor energy are conserved in the tubes, i.e. large sinks exist when the tubes (partially) pass an upstream turbine-rotor disk, such that the total tube cross-sectional area gradually shrinks until it reaches its originating turbine-rotor plane. At the originating rotor a large part of the remaining momentum/energy is removed, and due to effects of dissipation and further momentum/energy extraction at downstream turbines, the tubes rapidly shrink to zero. The additional one to two downstream tube sections that may be typically drawn before the tube disappears are not very enlightening; and therefore, only upstream sections are displayed for clarity.

Furthermore, the energy tubes in Figure~\ref{f:energytube} illustrate that the flux of total mechanical energy to the turbines in large wind-turbine-array boundary layers strongly depends on the stream-wise and span-wise spacing of turbines. In particular, when the span-wise spacing is large, energy is entrained from the sides, and only significantly further upstream is it entrained from above the turbines (see Fig.~\ref{f:energytube}d). In this case, the large span-wise spacing allows for high-speed flow to enter in between the turbine rows, where it further interacts sideways with the wake regions behind the turbines. For turbine arrays with narrow span-wise spacing (e.g. Fig.~\ref{f:energytube}c, and all staggered cases, Fig.~\ref{f:energytube}e--h), it is observed that the energy is entrained directly from the flow above, and less from the sides.

\begin{figure*}
  \begin{center}
  \includegraphics[width=0.45\textwidth]{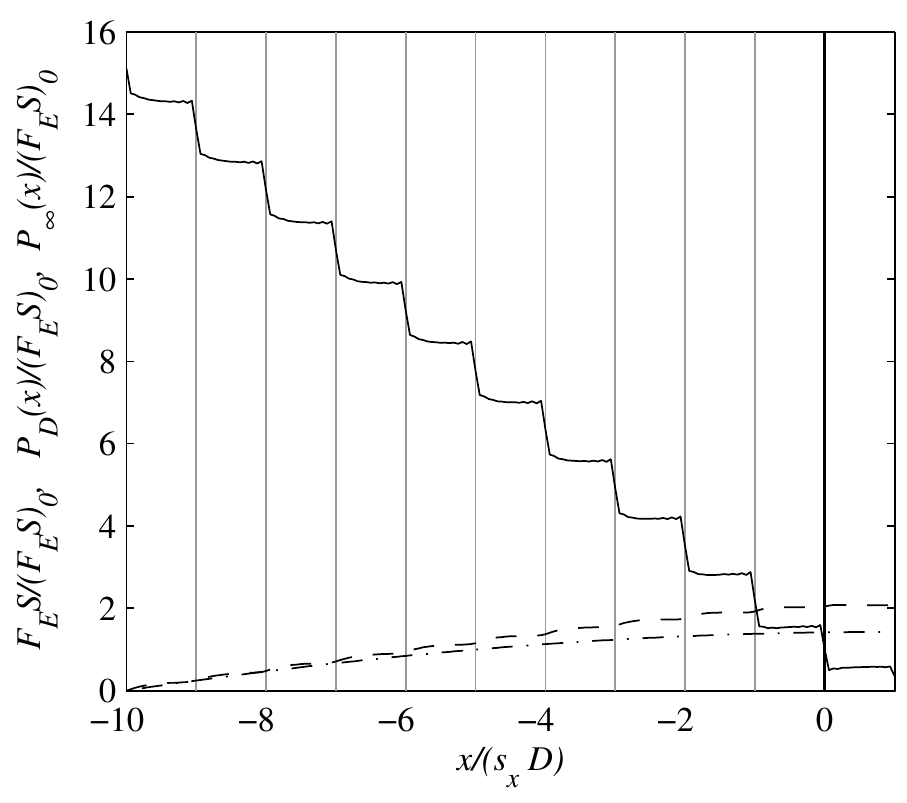}
  \end{center}
\caption{Analysis of mean-flow mechanical energy tubes for Case 1:  (---)  flux of total energy (normalized by target turbine value); ($--$) cumulative mean-flow dissipation by production of turbulence $P_D(x)$; and ($-\cdot$) cumulative power $P_\infty(x)$ inserted by the driving force $\nabla p_\infty$. }\label{f:tubeenthalpy}
\end{figure*}

By investigating the energy flux in the tube in Figure~\ref{f:tubeenthalpy}  (here for Case 1), it is seen that the energy level remains largely constant between turbine planes (apart from effects of production of turbulence, which appears to be significant only in regions  immediately downstream of the turbines, and power inserted by the driving pressure gradient), but drops significantly at the rotor disk locations as a result of the energy extraction by the turbines. This is quite different from the behavior of the stream tubes shown in  Figure~\ref{f:streamtubeeval}. We further find that results for Case 1 and Case 1R (not shown) are essentially the same (cf. also Appendix~\ref{s:windfarmsimulations}).

\begin{figure*}
  \begin{center}
  \includegraphics[width=0.45\textwidth]{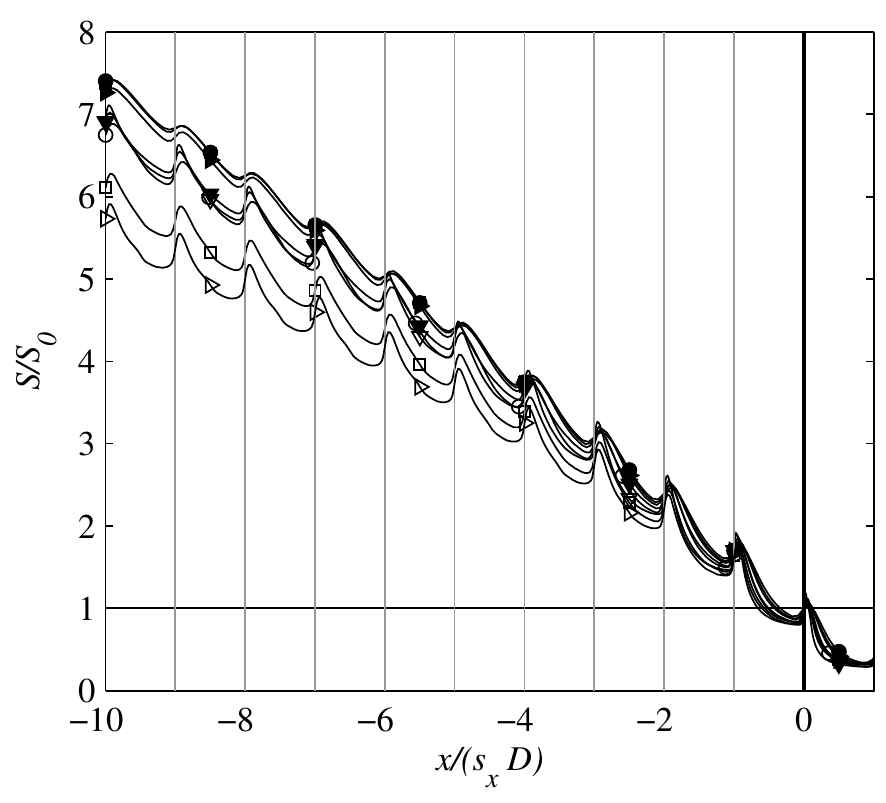}{\scriptsize (a)}
  \includegraphics[width=0.45\textwidth]{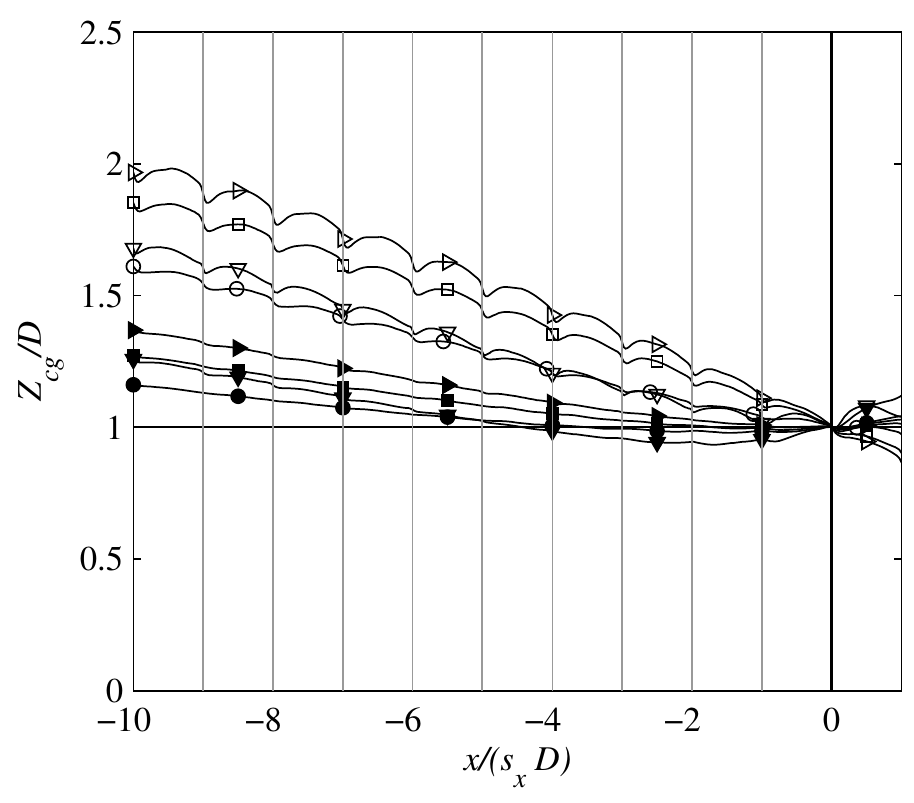}{\scriptsize (b)}
  \end{center}
\caption{(a) Evolution of cross-section area of energy tubes, and (b) geometric center $Z_{cg}$ of the energy tubes with closed symbols: aligned case; and open symbols: staggered case.  $\blacksquare, \square$: Case 1,  5; $\circ, \bullet$: Case 2, 6; $\blacktriangleright,\vartriangleright$: Case 3, 7; and $\blacktriangledown,\triangledown$: Case 4, 8.}\label{f:tubeenthalpyb}
\end{figure*}

Finally, in Figure~\ref{f:tubeenthalpyb}(a), the evolution of the energy-tube area is evaluated as function of upstream distance for the different wind-turbine arrays, while in Figure~\ref{f:tubeenthalpyb}(b) the tube geometric center is displayed. Some differences between aligned and staggered cases are observed. In particular, the geometric center of the tubes moves upward more appreciably for the staggered cases -- due to their narrower span-wise spacing, there is less room for sideways expansion.

\section{Discussion}\label{s:discussion}

The visualization of mechanical-energy tubes, as shown in Figure~\ref{f:energytube}, provides an intuitive understanding about the region of the flow supplying the power that ultimately is available at individual turbines. In this section, further features of the energy tubes are discussed.

First, it is useful to recall that the tube section associated with upstream plane $x/(s_xD) = -n$ (e.g., in Figure~\ref{f:energytube}) also corresponds to the section in the plane $x=0$ of a tube associated with a turbine at downstream distance $x/(s_xD) = n$. This is a direct consequence of the periodic nature of the flux vector field. Now, recall that changes in energy flux along the tubes are only due to internal sources and sinks, i.e. the power inserted by the driving force $\nabla p_\infty$ ($P_\infty$), the power extracted by turbines ($P_T$), and the production of turbulence ($P_D$). Hence, except for the relatively small difference between $P_D$ and $P_\infty$ compared to  $P_T$,  the sections in Figure~\ref{f:energytube} (omitting the minus signs in their numbering), may be roughly interpreted as containing the total flux of mechanical energy in the current cross section of the wind farm that will be extracted away in the next $n$ turbine rows.

\begin{figure}
  \begin{center}
  \includegraphics[height=0.5\textwidth]{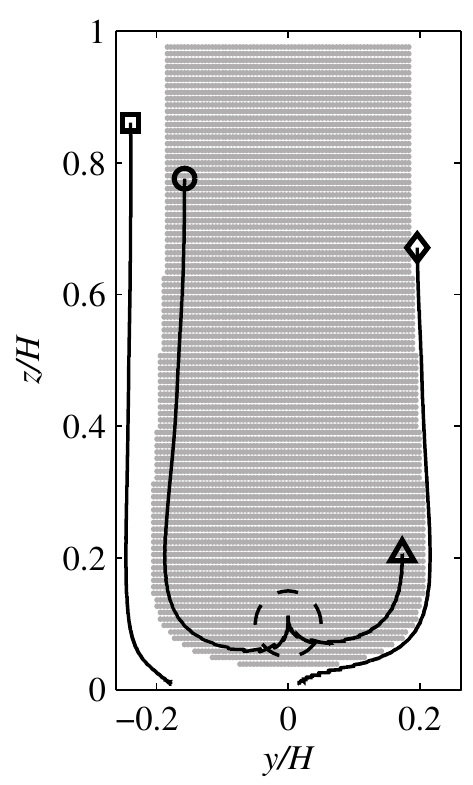}{\scriptsize(a)}
  \includegraphics[height=0.5\textwidth]{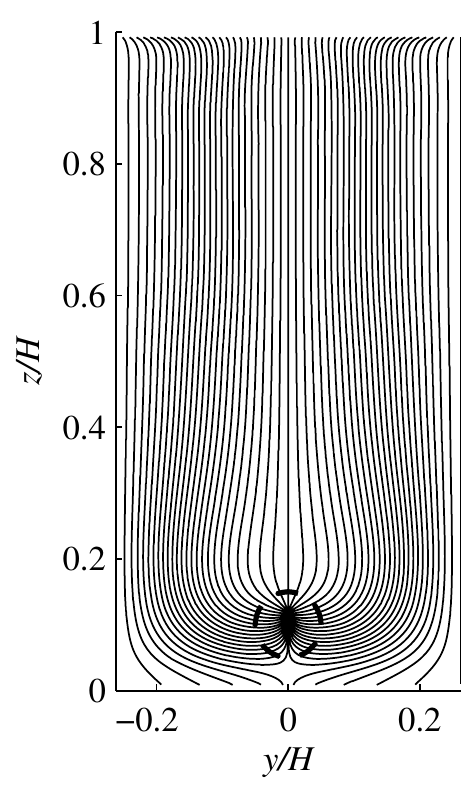}{\scriptsize (b)}
  \end{center}
\caption{(a) Domains of attraction of energy-flux lines through a cross section of the boundary layer for Case 1. White domain: flux lines are attracted to the ground surface; gray domain: flux lines pass through a turbine, and are attracted to a line close to the turbine center. Full line: four different flux lines and their projected trajectory (starting from points marked with $\square, \circ, \vartriangle, \lozenge$ respectively). (b) Projected flux lines in a turbine plane, obtained from the stream-wise averaged flux vector field $\langle\f{\vec{F}}_E\rangle_x$ (for Case 1).   }\label{f:tubeareaCase1}
\end{figure}

Furthermore, considering the subsequent sections of energy tubes in Figure~\ref{f:energytube}, we observe for all cases that the side and bottom boundaries of the sections asymptotically converge for increasing $n$. This suggests that the total boundary layer region is divided into a region covered by turbine tubes, and a region which is not. We investigate this in detail by seeding the cross section of the boundary layer with a large number of points, and constructing the downstream flux lines over a large number of periodic cycles. We first focus on Case 1 in Figure~\ref{f:tubeareaCase1}: 100 by 100  points are seeded on a Cartesian grid covering a cross section of the boundary layer. Subsequently flux lines are tracked over a large number of cycles to determine their attractors. Making connections to dynamical systems, we remark that the flow corresponding to the energy flux lines is not conservative, and so unlike the velocity field which does not possess attractors due to volume conservation, the energy flux lines can have attractors and repellers.  In Figure~\ref{f:tubeareaCase1}(a), the white (uncolored) area, represents all points from which the flux lines are attracted to the ground surface. The gray shaded area shows the points whose energy flux lines will eventually pass through a wind turbine disk. These flux lines are all attracted to what is seen as a point in the current cross (Poincar\'{e}) section. To illustrate representative trajectories of flux lines from different areas, we selected four points of which the flux lines (projected onto the turbine plane) are drawn. From this it is seen that points from white areas are indeed attracted by the ground surface, while points from the gray area end up in a flux line close to the turbine center.

In Figure~\ref{f:tubeareaCase1}(b), we show the energy flux lines from the 2-D vector field obtained from steam-wise averaging the energy-flux vector field according to $\langle {\vec{F}}_E\rangle_x$. This figure illustrates, in an integrated sense, how the energy flux lines are attracted to different regions in the boundary layer. Note that the comparison with Fig.~\ref{f:tubeareaCase1}(a) is qualitative only, and not exact: the lines in Figure~\ref{f:tubeareaCase1}(a) are obtained through Lagrangian tracking. The effective flux field is changing in $x$ over one period due to the expansion of the flow close to the turbines -- these effects are however relatively small, such that a comparison remains insightful.

\begin{figure}
  \begin{center}
  \includegraphics[height=0.5\textwidth]{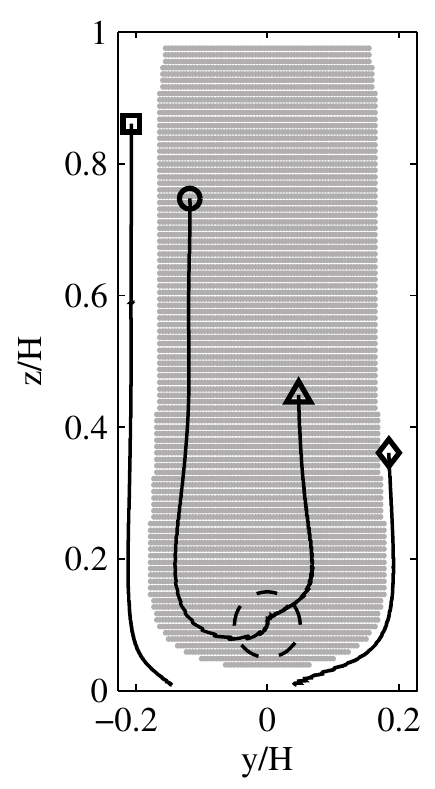}{\scriptsize (c)}
  \includegraphics[height=0.5\textwidth]{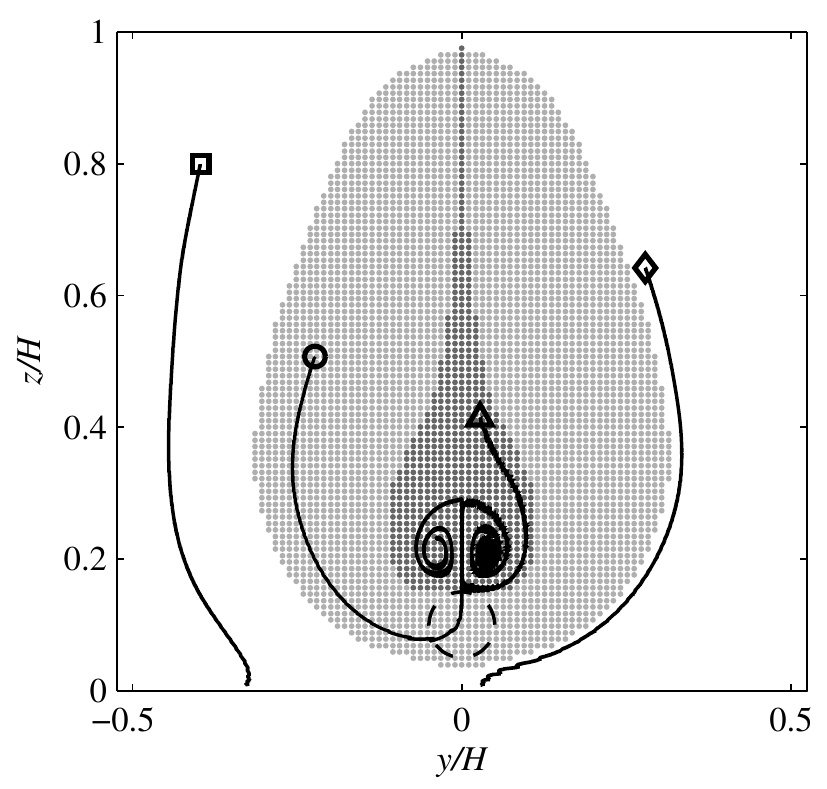}{\scriptsize (d)}
  \end{center}
\caption{Domains (basins) of attraction of energy-flux lines through a cross section of the boundary layer. (a) Case 3 -- White domain: flux lines are attracted to the ground surface; gray domain: flux lines pass through a turbine, and are attracted to a line close to the turbine center. (b) Case 4 --  White domain: flux lines are attracted to the ground surface; gray domains: flux lines are attracted into a spiral, either on top left or top right of the turbine; light gray domain: flux lines pass though turbine disk; dark gray: flux lines do not pass through turbine disk. Full line: four different flux lines and their projected trajectory (starting from points marked with $\square, \circ, \vartriangle, \lozenge$ respectively). }\label{f:tubearea}
\end{figure}

Basins of attraction for most other cases are similar to Case 1. For instance, in Figure~\ref{f:tubearea}(a) the basins of attraction for Case 3 are shown. As for Case 1, a relatively large part of the flow total energy is ``attracted'' towards the turbine, while in-between turbine rows, a smaller area is attracted to the ground (but dissipated before reaching it since no work is being performed at the bottom boundary). We find one case where this picture differs significantly, i.e. Case 4 with $s_x=15.7$ and $s_y=10.5$, which is shown in Figure~\ref{f:tubearea}(b). This is the case with the widest span-wise turbine spacing (and together with Case 8 has the highest average turbine spacing considered). Three distinct domains are now observed: a domain where flux lines are attracted to the ground (in white), and two gray domains where the flux lines are attracted along spirals to the top left and top right of the turbine row. Finally, flux lines in the lighter gray domain first pass through a turbine disk, where energy is extracted. Flux lines in the darker gray domain do not pass through a turbine. The result in Figure~\ref{f:tubearea}(b) is interesting, as it suggests that the available driving power for this case is not maximally used for wind-energy conversion. In large domains of the boundary layer (i.e. the white and dark-gray domains), the driving power is balanced by dissipation through production of turbulence, and only in the light-gray area is driving power converted to useful energy.

\section{Conclusions}\label{s:conclusions}
In the current work we explore the concept of momentum and energy transport tubes to study the three-dimensional mean fluxes of these properties in turbulent flows. These tubes are constructed based on transport vector fields, which include effects of Reynolds stresses and mean-flow viscous stresses. In particular, when transport processes are dominated by Reynolds stresses and turbulence instead of mean-flow convection, such transport tubes are an interesting means of visualizing where the momentum or the energy in the flow originate and/or are transported to.

As illustrative application,  we study stream-wise axial momentum and energy tubes in fully developed wind-turbine array boundary layers. Analyzing conventional stream tubes, we find that on average, the mean flow volume (mass) passing through a wind turbine disk comes from below the turbine, and is downstream ejected into layers above the turbines. Based on the energy tubes, we find that the energy takes a different path to reach the wind turbine locations.  Depending on turbine arrangement, there are two distinct paths and mechanisms taken by the  energy as it reaches the turbines: a sideways flux and a top-down flux. Sideways fluxes themselves are fed by a top-down flux in regions outside the turbine wake area. For large span-wise turbine spacings, sideways fluxes of energy  dominate; for small span-wise spacings, only the top-down mechanism is dominant.

Further investigating total-mechanical energy tubes and flux paths, we find that they define different basins of attraction in the boundary layer. In one part of the boundary layer, energy flux lines are attracted to the ground surface, while around and above turbines, flux lines are typically attracted to the turbine disk region. In some cases, attracting points above the wind turbine region were also observed. The relative size of these domains of attraction depend on turbine spacing and arrangement pattern.

In closing, we point out that it also may be interesting to consider momentum and energy flux lines and tubes in the case of laminar flows, where only the viscous fluxes provide differences to regular velocity and stream-lines  (some simple examples were provided in Section~\ref{s:laminar}). Moreover, there exists the possibility of fruitful analogies between physical-space trajectories of the generalized transport vector field and trajectories in phase-space of dissipative dynamical systems, similarly to how Hamiltonian dynamical systems provide useful analogies for laminar-flow chaotic mixing \citep{Aref1984,Ottino1989}.

C.M. acknowledges the National Science Foundation for support (grant \# NSF-CBET 1133800). J.M. acknowledges the Flemish Science Foundation for support (grant \# G.0376.12). The authors further thank Jos\'{e} Lebron, Luciano Castillo and Jonas Boschung for fruitful conversations and comments. Large-eddy simulations were performed on the computing infrastructure of the VSC  -- Flemish Supercomputer Center, funded by the Hercules Foundation and the Flemish Government.

\appendix

\section{Wind-farm simulations: governing equations and computational setup}\label{s:windfarmsimulations}
This appendix provides additional details about the LES. The methodology has already been presented in \cite{Calaf2010}. Moreover Case 1--4 (see Table~\ref{tab:cases}) in the current work correspond with Cases A3, K, J, and G of \cite{Calaf2010}, respectively.

The current work visualizes the interactions between an infinite wind farm and a neutral boundary layer. We simulate rough-wall fully-developed boundary-layers in a periodic domain, driven by a pressure gradient. We do not include Coriolis forces. While these are present in the atmosphere at the larger scales of the flow, the main rationale underlying this approach (see discussion in \cite{Calaf2010}) is based on the classical hypothesis that inner-layer dynamics of a boundary layer ($y<0.15H$) are approximately independent of outer layer effects. We presume that turbines (with height $\approx 100$m) are situated in the inner layer of the boundary layer, which is relevant in practice for many atmospheric cases with boundary layer depths above 1 km. This approach was used before by \cite{Calaf2010}, where it allowed for the characterization of increased surface roughness induced by a wind farms, and the derivation of algebraic surface-roughness models. Later, wind-farm performance obtained based on these surface-roughness models were shown to compare well with observations in the Horns Rev, and Nysted wind farms \citep{Meyers2012}.

\subsection{Large-eddy simulations}
We consider thermally  neutral flow that is driven by an imposed pressure gradient. LES solves the the filtered incompressible Navier–-Stokes equations for neutral flows and the continuity equation, i.e.,
\begin{eqnarray}
\Pd[\ft{u}_i]{x_i} &=& 0 \\
\Pd[\ft{u}_i]{t} + \Pd[\ft{u}_i \ft{u}_j]{x_j} &=& -\frac{1}{\rho} \Pd[\ft{p}]{x_i} + \Pd[\tau_{ij}]{x_j} + f_i, \label{e:NS}
\end{eqnarray}
where $\ft{u}_i$ is the resolved velocity field, $\ft{p}$ the pressure, $\tau_{ij}$ are the subgrid-scale stresses, and where the density $\rho$ is assumed to remain constant. Furthermore, $f_i$ represents forces introduced by the turbines on the flow (see discussion below). Since the Reynolds number in atmospheric boundary layers away from the bottom boundary is very high, we neglect the resolved effects of viscous stresses in the LES. The deviatoric part of the subgrid-scale stresses is modeled here with the conventional \cite{Smagorinsky1963} model, with a constant coefficient Cs=0.14 (the trace of the subgrid-scale stresses $\tau_{kk} /3$ is not modeled, but instead absorbed into the pressure term, as is common practice in LES of incompressible flow). Near the bottom surface, the Smagorinsky length scale $\lambda$ ($=C_s\Delta$ far from the surface) is damped using the classic wall damping function of \cite{Mason1992}, i.e. $\lambda^{-n} = [C_s\Delta]^{-n} + [\kappa(z+z_{0,lo})]^{-n}$, where we take $n=3$. Other works have also used more
advanced subgrid-scale models in LES of wind farms (e.g. the scale-dependent Lagrangian model of  \citealp*{BouZeid2005} was used for several of the simulations presented in \cite{Calaf2010}), but the differences in mean velocity and Reynolds stress distributions were found small, especially in regions where the transport tubes of interest in this study are mostly located.

In the stream-wise and span-wise directions, we use  periodic boundary conditions. The top boundary uses zero vertical velocity and zero shear stress condition. At the bottom surface, we impose zero normal velocity and use a classic, imposed wall-stress boundary condition. It  relates the wall stress to the velocity at the first grid-point using the standard log (Monin--Obukhov) similarity law \citep{Moeng1984}:
\begin{equation} \label{eq:wallstress}
 \tau_{w1}=-\left(\frac{\kappa}{\ln{(z/z_{0,lo})}}
\right)^2 \left( \widehat{\widetilde{u}}^2+
\widehat{\widetilde{v}}^2\right)^{0.5} \widehat{\widetilde{u}}
\end{equation}
\begin{equation}
\tau_{w2}=-\left(\frac{\kappa}{\ln{(z/z_{0,lo})} } \right)^2 \left(
\widehat{\widetilde{u}}^2+ \widehat{\widetilde{v}}^2
\right)^{0.5} \widehat{\widetilde{v}},
\end{equation}
where the hat on $\widehat{\widetilde{u}}$ and $\widehat{\widetilde{v}}$ represents a local average obtained by
filtering the LES velocity field with filter width $4\Delta$ (see \cite{BouZeid2005} for more details about such filtering).

The simulation code uses a pseudo-spectral discretization in the horizontal directions. The nonlinear convective terms and the SGS stress are de-aliased using the 3/2 rule \citep{can88}. Message Passing Interface (MPI) is used to run the simulations in parallel mode, and the FFTW library is employed for Fourier transforms \citep{FFTW05}. In the vertical direction, a fourth-order energy-conservative finite-difference discretization is used \citep{ver03}. Time-integration is performed using a classical four-stage fourth-order Runge--Kutta scheme.

\subsection{Turbine forces}
For the wind-turbine forces, we use an actuator disk model. These type of turbine representations were adopted in LES by \cite{Jimenez2007,Jimenez2008,Ivanell2009,Calaf2010} amongst others. Recently, they were thoroughly validated against wind-tunnel data by \cite{Wu2011}. Details of the current implementation are given in \cite{Meyers2010}. In this model, the thrust and tangential forces in the turbine rotor disk per unit actuator-disk area are given by
\begin{eqnarray}
F_t = -  \rho \frac{1}{2}\,C'_T\,\langle {\f{u}}^T \rangle_{d}^2,\\
F_\theta(r) = \frac12 C_P' \langle {\f{u}}^T \rangle_{d}^2 \frac{\langle \f{u} \rangle_{d}}{\Omega r} \label{e:Ftheta}
\end{eqnarray}
with the subscript `$d$' denoting an averaging over the turbine disk region, and the superscript `$T$' denoting time filtering or averaging over a time-scale of order $T$. Thus $\langle {\f{u}}^T \rangle_{d}$ is the  disk averaged and time-filtered velocity (further discussed below). The parameters $C_T'$ and $C_P'$ are modified thrust and power coefficients, defined based on the turbine disk velocity instead of the undisturbed upstream velocity as conventionally used. Their values are directly related to the aerodynamic lift and drag coefficients of the turbine blades, the blade geometry, etc. \citep{Meyers2010}. We use values of $C_T'=4/3$, and for the rotating (ADMR) case $C_P'=1$ (for the non-rotating case, the tangential forces are zero). For a lone-standing turbine, these values would correspond to conventional thrust and power coefficients of $C_T=0.75$ and $C_P \approx 0.42$ \citep{Meyers2010}. Finally, $\Omega$ is the turbine angular velocity, and $0<r<D/2$ the radial location on the turbine disk. For the ADMR case in the current work (Case1R), we use $\Omega R/u_* = 60$, which roughly corresponds to a tip-speed ratio of $\lambda \approx 6.7$ (based on the average boundary-layer velocity at turbine hub height).

To implement the forces $f_i$ in (\ref{e:NS}), the turbine forces $F_i$ (with axial, and tangential components $F_t$, and $F_\theta$) are first described in the turbine-rotor plane. In a second step, these forces are filtered using a Gaussian convolution filter on locations which correspond with the coordinates of the LES grid. We use a Gaussian filter, with filter width $\Delta=1.5 \Delta$ (and $\Delta$ is the grid spacing) to avoid Gibbs oscillations on the LES grid. A similar smoothing approach was used for actuator-line representations in \cite{Sorensen2002}. To evaluate the disk averaged local velocity $\langle {\tilde{u}} \rangle_{d}$ needed for the determination of the force $F_t$ and $F_\theta$, we employ the geometrical rotor footprint on the LES grid as a weighting function for the averaging. Moreover, $\langle {\f{u}}^T \rangle_{d}$ is obtained from $\langle \tilde{u} \rangle_{d}$ by using a one-sided exponential time-filter, using a time window of $Tu_*/H= 0.6$. Further details are found in \cite{Calaf2010} and \cite{Meyers2010}.

\subsection{Effects of numerical discretization}

\begin{figure*}
  \begin{center}
  \includegraphics[scale=0.6]{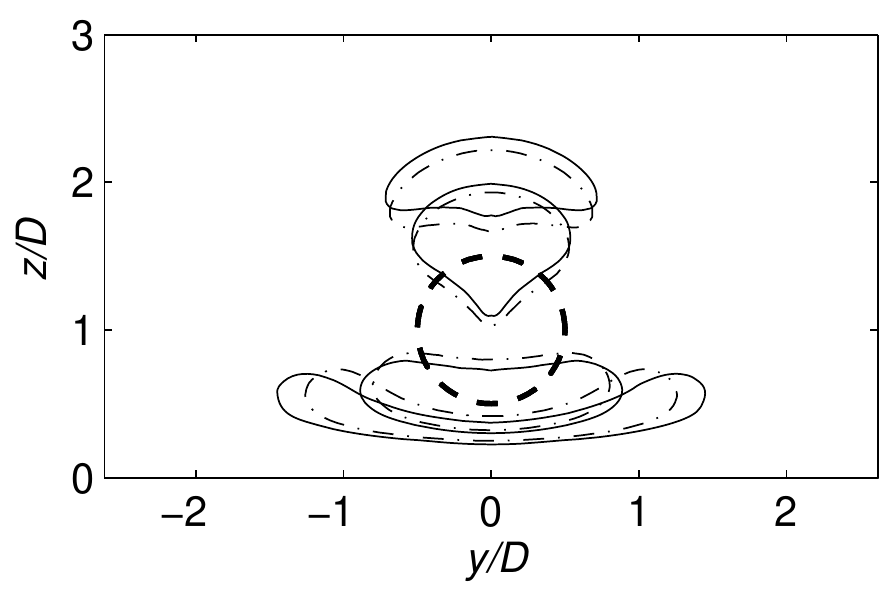}{\scriptsize (a)}
  \includegraphics[scale=0.6]{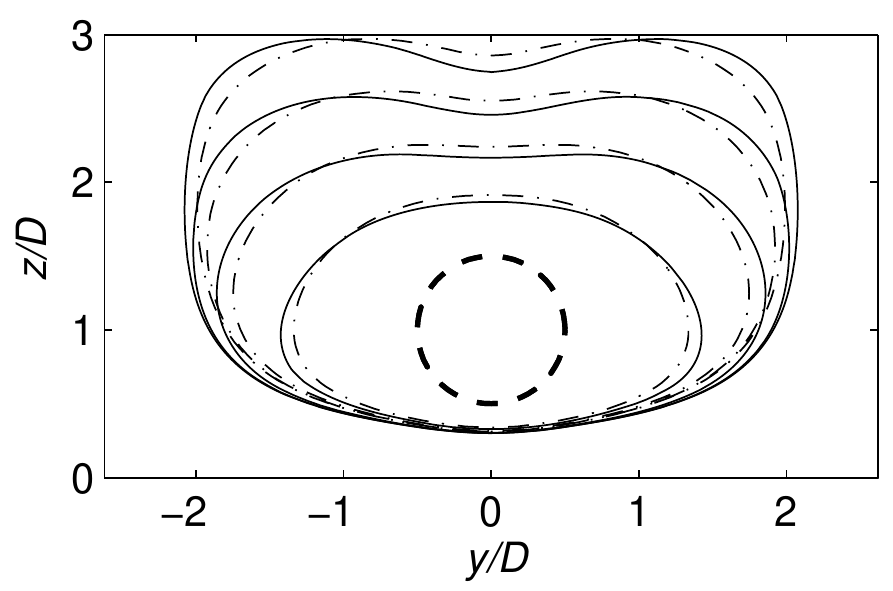}{\scriptsize (b)}
  \end{center}
\caption{(a) Upstream and downstream sections of stream tubes (at $x=\pm 5 s_xD$ and $\pm 10 s_xD$) and (b) upstream sections of total mechanical energy tubes (at $x=- n s_xD$ with $n= 5,10,15,20$) for (---): Case~1; and ($-\cdot$): Case~1F  --- see Table~\ref{tab:cases} for details.}\label{f:streamtubegridref}
\end{figure*}

All cases except Case~1F are discretized using similar domain and grid sizes. For the steam-wise and span-wise domain size we use $L_x= 2\pi H$ and $L_y=\pi H$ (with H the boundary-layer depth and domain height) to allow for the large scale turbulent structures that typically emerge in the boundary layers to be represented properly. The cell size in the simulations correspond to $\Delta y\approx \Delta z \approx 0.016H$, and $\Delta x \approx 0.05H$, except for Case~1F, where a finer mesh is used, with $\Delta y\approx \Delta z \approx 0.01H$, and $\Delta x \approx 0.033H$. The effect of refining the mesh on stream tubes and mechanical energy tubes is shown in Figure~\ref{f:streamtubegridref}. We observe that the shape of the tubes remains relatively unaffected, especially considering that the tube cross-sections shown correspond to very long integration distances of the stream and energy lines and small differences in mean velocity arising from the LES can accumulate while integrating the trajectories.

\subsection{Effects of wake rotation}

\begin{figure*}
  \begin{center}
  \includegraphics[scale=0.6]{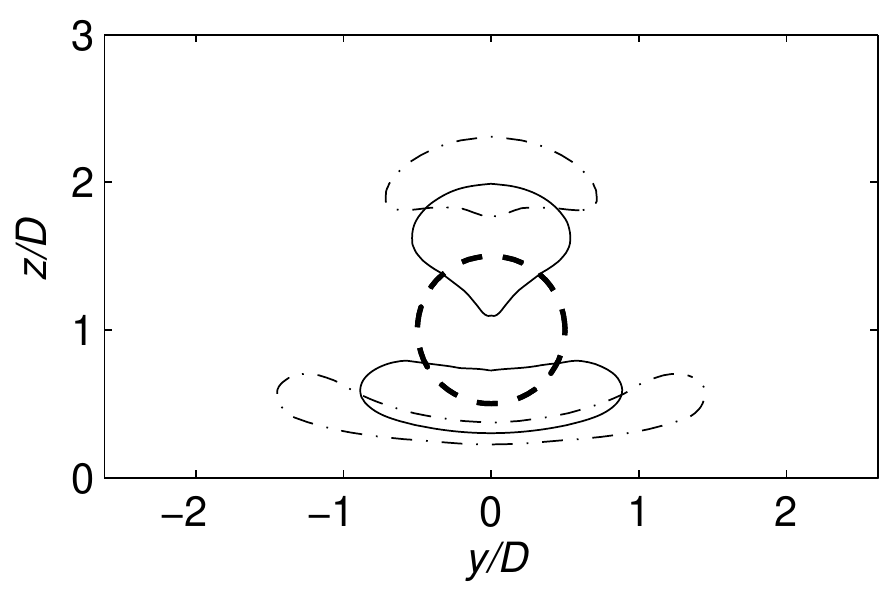}{\scriptsize (a)}
  \includegraphics[scale=0.6]{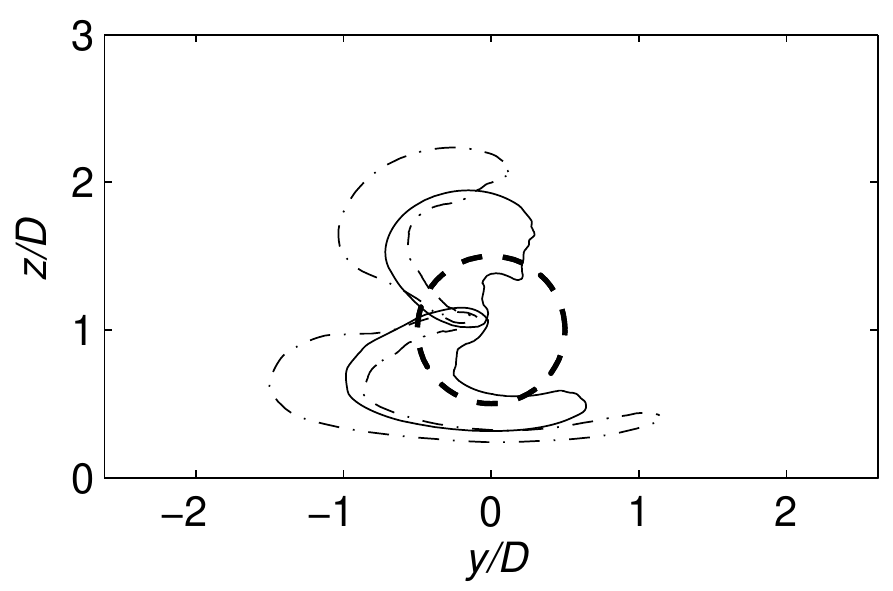}{\scriptsize (b)}
  \end{center}
\caption{Upstream and downstream sections of stream tubes for (a) non-rotating case (Case 1); and (b) rotating case (Case 1R) --- see Table~\ref{tab:cases} for details. ($--$): turbine rotor; (---): sections at $x=\pm 5 s_x D$; and ($-\cdot$): sections at $x=\pm 10 s_x D$.}\label{f:streamtubebis}
\end{figure*}

\begin{figure*}
  \begin{center}
  \includegraphics[scale=0.6]{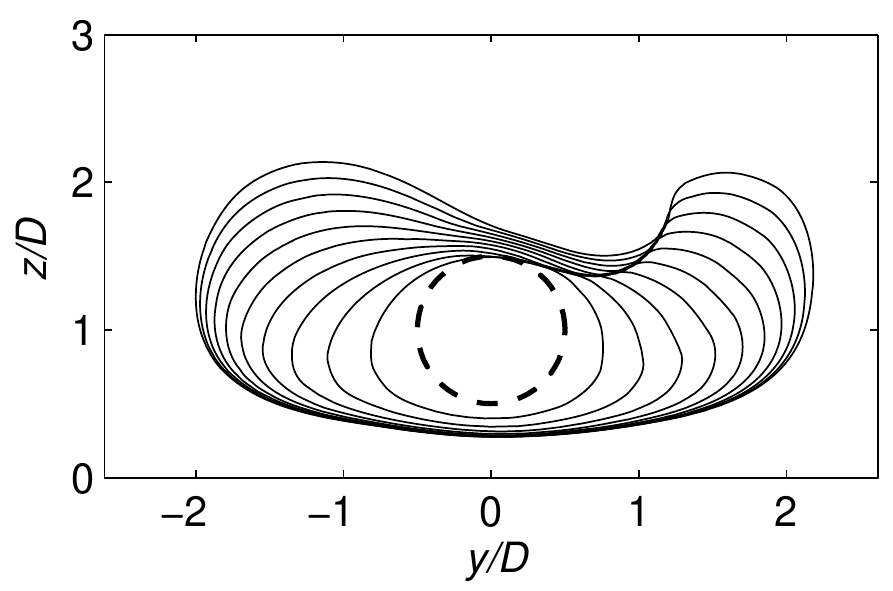}{\scriptsize (a)}
  \includegraphics[scale=0.6]{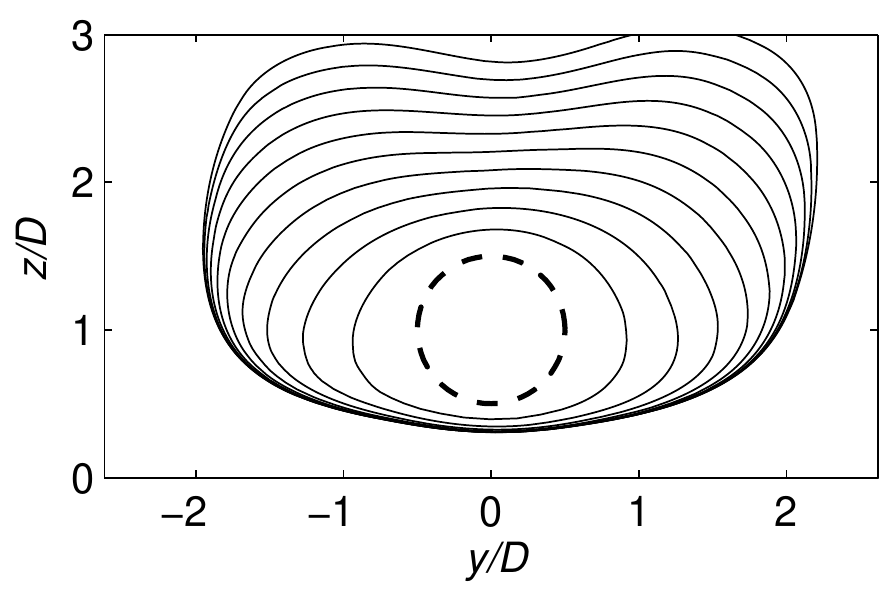}{\scriptsize (b)}
  \end{center}
\caption{Upstream sections of (a) axial momentum, and (b) mean-flow mechanical energy transport tubes for Case 1R. ($--$): turbine rotor; (---): sections at different upstream locations, with upstream distances corresponding to $x=-n s_x D$, and $n=2,4,\cdots, 20$.}\label{f:rotatingtubes}
\end{figure*}

In Figure~\ref{f:streamtubebis} we examine effects of wake angular momentum introduced via tangential forces on the evolution of the stream tubes, by comparing Case~1 and Case~1R. As expected, the induced swirl is associated with a twist of the stream tubes in upstream and downstream directions when compared to the symmetric non-rotating case. Nevertheless, also in the rotating case, mass is ejected upwards downstream from the turbines, while upstream it is entrained from below the turbines. Moreover, as already briefly discussed in Section~\ref{s:masstubes}, momentum fluxes and cross sectional area along the tube are not much influenced by wake rotation.

In Figure~\ref{f:rotatingtubes} the momentum and total mechanical energy tubes are displayed for Case 1R. When comparing with the non-rotating cases, i.e. Figure~\ref{f:momentumtube}(a) momentum, and Figure~\ref{f:energytube}(a) for mean-flow mechanical energy, it is appreciated that the effect of rotation appears less visible than for the conventional stream tubes (cf. Figure~\ref{f:streamtubebis}). In particular,  the effect of rotation is quite small for energy tubes. This is not unexpected, as modern wind farms operate at relatively high tip-speed ratios, leading to low torque (for a given amount of power) and low associated tangential forces,  such that effects on the Reynolds stresses, which are largely responsible for transport of energy, remain small.

\bibliographystyle{jfm}



\end{document}